\documentclass[journal,draftcls,onecolumn]{IEEEtran}
\usepackage{amsmath, amssymb, amsthm}

\usepackage{graphicx} 

\usepackage{enumerate}
\usepackage{multirow,array}
\usepackage{cite}
\usepackage{graphicx}
\usepackage{psfrag}
\usepackage{url}
\usepackage{amsmath}
\usepackage{amsthm}
\usepackage{array}
\usepackage{amssymb}
\usepackage{amsfonts}
\usepackage{float}
\usepackage{mathtools}

\usepackage{algorithm}
\usepackage{algpseudocode}

\usepackage{hyperref}
\usepackage{mathbbol}
\usepackage{standalone}
\usepackage{tikz}
\usetikzlibrary{arrows.meta,positioning,calc}

\hypersetup{
    colorlinks,
    citecolor=red,
    filecolor=black,
    linkcolor=blue,
    urlcolor=black
}

\newcommand{\F}{\mathbb{F}_q}

\newcommand{\CZ}{\mathcal{C}_{\mathrm{Z}}}

\newcommand{\LP}{\mathrm{LP}}

\DeclareMathOperator{\im}{im}

\newcommand{\R}{R_\ell}
\newcommand{\Rb}{R_{(b)}}
\newcommand{\wt}[1]{\lvert #1\rvert}
\newcommand{\wtH}[1]{\lvert #1\rvert_{\mathrm{H}}}
\newcommand{\ann}{\operatorname{ann}}
\newcommand{\dist}{\operatorname{dist}}
\newcommand{\CXp}{\mathcal{C}_X^{\perp}}
\newcommand{\Cb}{\mathcal{C}_b}

\usepackage{url}
\usepackage{ifthen}
\usepackage{cite}
\usepackage{longtable}
\usepackage{makecell}
\usepackage{algorithm}
\usepackage{color}
\usepackage{amsmath,amssymb,amsthm}
\usepackage{mathtools,bbm}  
\usepackage{graphicx}                        
\usepackage{xifthen,xparse}
\usepackage{float}
\usepackage{hhline}
\usepackage{physics}
\usepackage{stmaryrd}
\usepackage[table]{xcolor}
\usepackage{textcomp}
\usepackage{esvect}
\usepackage{algpseudocode}
\usepackage[T1]{fontenc}
\usepackage[caption=false,font=footnotesize]{subfig}

\allowdisplaybreaks

\newtheorem{theorem}{Theorem}

\newtheorem{corollary}[theorem]{Corollary}
\newtheorem{remark}[theorem]{Remark}
\newtheorem{lemma}[theorem]{Lemma}
\newtheorem{definition}[theorem]{Definition}

\theoremstyle{remark}
\newtheorem*{remark*}{Remark}
\newtheorem*{note*}{Note}

\NewDocumentCommand\dketbra{+m+g}{%
  \IfNoValueTF{#2}
    {\left\lvert #1 \right\rangle \left\langle #1 \right\vert}
  {\left\lvert #1 \right\rangle \left\langle #2 \right\rvert}%
}

\NewDocumentCommand\dbraket{+m+g}{%
  \IfNoValueTF{#2}
    {\left\langle #1 \vert #1 \right\rangle}
  {\left\langle #1 \vert #2 \right\rangle}%
}

\global\long\def\Pr{\mathbb{P}}

\newtheorem{proposition}[theorem]{Proposition}

\begin{document}


\title{Quasi-Cyclic Qudit Lifted Product Codes with \\ Almost Linear Minimum Distance}

\author{
Shantom~K.~Borah, 
Abhinav~Vaishya, 
V. ~Lalitha,~\IEEEmembership{Senior~Member,~IEEE,} 
Asit~Kumar~Pradhan, 
Bane~Vasi\'{c},~\IEEEmembership{Fellow,~IEEE,} 
and~Narayanan~Rengaswamy,~\IEEEmembership{Senior~Member,~IEEE}%

\thanks{V.~Lalitha is with the International Institute of Information Technology, Hyderabad 500032, India; e-mail: lalitha.v@iiit.ac.in.}
\thanks{A.\ Vaishya is with Departamento de Matemática, Instituto Superior Técnico, Universidade de Lisboa, 1049-001 Lisboa, Portugal, and Instituto de Telecomunicações, 1049-001 Lisboa, Portugal. Part of this work was completed while he was with the Indian Institute of Science, Bengaluru, India; e-mail: abhinav.vaishya@tecnico.ulisboa.pt}
\thanks{S.~K.~Borah, A.~K.~Pradhan, B.~Vasi\'{c} and N.~Rengaswamy are with the Department of Electrical and Computer Engineering at the University of Arizona, Tucson, AZ 85721, USA; e-mail: \{ shantomborah, asitpradhan, narayananr \}@arizona.edu.}
}





\maketitle

\begin{abstract}
    Quantum Low-Density Parity-Check (QLDPC) codes are a promising approach towards fault-tolerant quantum computation with low space and time overhead. Qudit systems offer several advantages over qubits, including a more efficient information representation, native support for mixed-dimensional problems, improved magic state distillation protocols, and a richer mathematical structure for code design. Asymptotically good QLDPC codes have been constructed via the lifted product (LP) method using lifts over non-abelian groups, and there is strong evidence that abelian lifts cannot achieve a linear minimum distance. Cyclic lifts are nevertheless attractive in practice, since they produce quasi-cyclic codes whose circulant structure is exploited by hardware decoder implementations, a consideration that is especially important for non-binary decoders with their costly check node operations. In this work, we introduce a qudit generalization of the LP construction over cyclic group algebras, which extends hypergraph product and lifted product codes to arbitrary prime-power qudit dimensions and yields quasi-cyclic qudit CSS codes for every stacking order. We provide a detailed analysis of qudit LP codes and explore their connection to additive quantum codes to establish bounds on encoding rates. We further show that, in the fully stacked case, qudit LP codes built from lifted Tanner codes on expander graphs form families with $\Theta(\log N)$ logical qudits and an almost linear minimum distance $\Theta(N/\log N)$, and we prove a minimum distance bound that holds for a general class of lifted product factors. In addition, we study decoding strategies for qudit QLDPC codes. Specifically, we discuss an additive decoder suitable for general qudit codes as well as a linear decoder optimized for fully stacked constructions. Monte Carlo simulations under a depolarizing channel with equal symbol error probability show that qudit codes achieve lower frame error rates than comparable qubit codes and that the linear decoder provides further performance gains for fully stacked codes. Our results provide families of quasi-cyclic qudit QLDPC codes with almost linear minimum distance, along with concrete benchmarks against message-passing decoders based on non-binary belief propagation.
\end{abstract}


\section{Introduction}
\label{sec:intro}
\IEEEPARstart{T}{he} field of quantum information processing represents a fundamental paradigm shift in the manner in which we understand the operation of modern information processing technologies. Unlike classical information processing systems, which are often described as state machines operating over a finite set of states, quantum systems leverage inherently quantum phenomena such as superposition and entanglement to unlock new regimes of operation. As early as the early 1980s, Feynman~\cite{feynman1982simulating} speculated that an information processing system relying on quantum phenomena could potentially hold the key to significant speedups to many problems of classical interest. Indeed, this speculation was largely driven by the considerable difficulties faced by physicists of that era in simulating the behavior of quantum systems. Several decades later, we now live in an era where these abstract theoretical ideas are beginning to take concrete forms. While modern quantum computing systems are still not sufficiently evolved to be able to provide a concrete quantum advantage in relation to a problem of commercial utility, the progress in both quantum hardware and software in the last few decades has been spectacular~\cite{google2025willow, bluvstein2026faulttolerant}, with recent breakthroughs in error correction and qubit stability demonstrating that the fundamental hardware building blocks are now in place, marking a decisive shift from theoretical concepts to experimental reality. Central to the vision of a universal fault-tolerant quantum computer is the field of Quantum Error Correction (QEC), which concerns itself with the detection and correction of errors in quantum information without directly measuring the encoded quantum state.

Within the field of QEC, the use of Quantum Low-Density Parity-Check (QLDPC) codes has been widely recognized as one of the most promising paths to fault-tolerance~\cite{breuckmann2021quantum, vasic2026quantum}. QLDPC codes, similar to their classical LDPC counterparts, are characterized by low weight parity-check matrices. Early work on QLDPC codes dates back to the work of MacKay et al.~\cite{mackay2004sparse}, with the introduction of the bicycle, unicycle and multi-cycle code constructions. The hypergraph product construction of Tillich and Z\'{e}mor~\cite{tillich2013quantum} was the next major milestone in the development of QLDPC codes, where the authors were able to show that the hypergraph product construction admitted QLDPC codes with a linear scaling of the code dimension and a square root scaling of the minimum distance with respect to block length. This was on par with the known code parameters for the surface and toric codes~\cite{dennis2002topological}, which represented the state-of-the-art in QEC at that time. Subsequently, several QLDPC code constructions were published~\cite{hastings2021fiber, breuckmann2021balanced} that slowly pushed this boundary upwards. Eventually, this line of research culminated in the work of Panteleev and Kalachev, who first used the lifted product (LP) construction to obtain QLDPC codes with an almost linear minimum distance~\cite{panteleev2021quantum}, and subsequently demonstrated a linear scaling of both code dimension and minimum distance with respect to block length~\cite{panteleev2022asymptotically}. At around the same time, two other constructions, the quantum Tanner construction~\cite{leverrier2022quantum} and the DHLV construction~\cite{dinur2023good}, also achieved a similar linear scaling of code parameters with respect to block length. This flurry of breakthrough results in 2021--2023 prompted a significant increase in research activity into other aspects of implementing QLDPC codes for QEC, such as performance and decoding under circuit-level noise~\cite{maan2026decoding, pacenti2025turbo}, logical operations~\cite{ide2025fault, malcolm2026computing}, magic state factories~\cite{menon2026singleshot}, etc. Today, QLDPC codes are considered by many to be the state-of-the-art approach to QEC.

In parallel to the above theoretical advances, the field of quantum computing has also seen similar breakthroughs on the hardware front over the past two decades~\cite{google2025willow}. Today, quantum hardware can be realized in the form of several hardware modalities including, but not limited to, superconducting qubits~\cite{blais2021circuit}, photonic systems~\cite{slussarenko2019photonic}, trapped-ion systems~\cite{bruzewicz2019trapped}, neutral atoms~\cite{saffman2010quantum}, and topological qubits~\cite{nayak2008non}. From a theoretical perspective, most of these hardware platforms can be modeled as a finite dimensional Hilbert space with some dimension $q$. While some of these systems admit an infinite dimensional Hilbert space (such as, for instance, superconducting and photonic systems), engineering constraints often require us to designate a chosen finite-dimensional subspace of such a system as the computational space. In the typical case of qubit-based quantum computing, the chosen dimensionality of this computational subspace is typically $2$. However, the fact that the underlying physical hardware does, in fact, support multi-level quantum systems (also known as qudits) in most cases raises the question of whether this extra degree of freedom can be meaningfully utilized for improved system design. Indeed, recent work on magic state distillation has established significant reductions in overhead requirements when operating with qudits rather than qubits~\cite{prakash2025low, cervia2025magicstate}. Moreover, from a classical coding theory perspective, non-binary message passing decoders for classical LDPC codes are known to perform significantly better than their binary cousins~\cite{davey1998low, declercq2007decoding}. This motivates the question of whether such gains may be adapted to the quantum case as well.

Some of the earliest works on qudit error correction were done by Ashikhmin et al.~\cite{ashikhmin2001nonbinary} and Ketkar et al.~\cite{ketkar2006nonbinary}. These two papers laid the early foundations of the qudit Pauli group over Galois fields which allowed classical non-binary error correcting codes to be used as quantum CSS codes for qudit systems. Ezerman et al.~\cite{ezerman2013css} subsequently extended the CSS construction to pairs of nested subfield-linear codes, and Nadkarni and Garani~\cite{nadkarni2021mathbb} showed that the generalization of the CSS framework to qudits of dimension $p^m$ yields two distinct classes of codes, namely $\mathbb{F}_p$-linear and $\mathbb{F}_{p^m}$-linear CSS codes, a distinction that underlies the stacked constructions studied in this paper. Nadkarni and Garani further developed a framework for qudit stabilizer codes over $\mathbb{F}_{p^m}$ based on classical additive codes, in which a qudit of dimension $p^m$ is decomposed into $m$ sub-qudits of dimension $p$~\cite{nadkarni2021nonbinary}, together with encoding procedures~\cite{nadkarni2018encoding, nadkarni2021encoding} and syndrome computation architectures~\cite{nadkarni2021quantum} for such codes; an overview of this line of work can be found in~\cite{garani2023theory}. In addition, several algebraic constructions have been proposed for non-binary error correction involving qudits, including non-binary quantum Reed-Muller codes \cite{sarvepalli2005nonbinary}, non-binary quantum Bose-Chaudhuri-Hocquenghem (BCH) codes \cite{la2014construction}, entanglement-assisted Reed--Solomon codes over qudits \cite{nadkarni2021entanglement}, and graphical non-binary quantum codes \cite{hu2008graphical}. Despite this early progress, research into the use of non-binary QLDPC codes for qudit error correction has been relatively sparse. Recently, the hypergraph product construction was generalized to the qudit case by Borah et al.~\cite{borah2024nonbinary} and subsequently Spencer et al.~\cite{spencer2026qudit} have proposed qudit generalizations of several other families of QLDPC constructions. The constructions proposed in both of these works, however, stop short of a generalization of the lifted product construction to the qudit case.

As discussed in the preceding paragraphs, the LP construction has played a critical role in the domain of qubit error correction by providing the first construction that admits an asymptotically good scaling of code parameters with respect to block length. The LP construction itself is not tied to the binary field: Panteleev and Kalachev define it over a general group algebra $\F G$\footnote{Throughout this paper, we use $GF(q)$ and $\F$ interchangeably to denote the Galois field of order $q$.}, and their asymptotically good codes exist over every finite field $\F$~\cite[Thm.~2]{panteleev2022asymptotically}, which, in the language of this paper, corresponds to fully stacked qudit codes. Their construction, however, relies on lifts over non-abelian groups, and they observe that an upper bound on the minimum distance of codes obtained from abelian lifts provides strong evidence that non-abelian groups are necessary to obtain asymptotically good LP codes~\cite{panteleev2022asymptotically}. For abelian lifts, the best known scaling is the almost linear minimum distance $\Theta(N/\log N)$ of~\cite{panteleev2021quantum}. From a practical standpoint, however, cyclic lifts are highly desirable. A lift over the cyclic group yields a quasi-cyclic code, whose parity-check matrix consists of circulant blocks. Hardware decoder architectures for classical LDPC codes exploit this structure extensively, since the circulant blocks allow processing units and memory to be reused through cyclic shifts, which considerably reduces the decoder footprint~\cite{mansour2003high, zhang2016vlsi}. This consideration is even more important for non-binary codes, where each check node update operates on probability vectors over $GF(q)$ and is considerably more expensive than its binary counterpart~\cite{declercq2007decoding}, and quasi-cyclic structure is likewise central to efficient non-binary decoder architectures~\cite{zhang2011efficient}. In this work, we therefore adopt cyclic lifts as a design constraint, trading the asymptotically good scaling of non-abelian constructions for a quasi-cyclic structure, while retaining an almost linear minimum distance. We regard the latter as supporting the hypothesis that such codes can perform well at practical finite lengths, which we examine numerically in Sec.~\ref{sec:results}. We note that Panteleev and Kalachev also remark that the lifted product of two cyclically lifted Tanner codes yields quasi-cyclic codes with a linear number of logical qudits and minimum distance $\Theta(N/\log N)$~\cite{panteleev2022asymptotically}; incorporating such constructions into the framework developed here is left for future work.

It should be noted, however, that a qudit generalization of the LP code based on lifts of finite field elements has already appeared in the work of Panteleev and Kalachev (see Example 2 in~\cite{panteleev2021quantum}). However, it was shown in~\cite{borah2024nonbinary} that this approach to the qudit LP construction reduces to the stacked hypergraph product construction over a finite field. Hence, such an approach inherits the limitations of the hypergraph product and does not benefit from the improved distance scaling of lifted products. The construction proposed in this work is based on the use of lifts over $\F$-algebras rather than lifts over the finite field elements themselves, and does not suffer from this limitation.

This paper provides a comprehensive analysis of both the qudit hypergraph product and the qudit lifted product constructions, a family of quasi-cyclic qudit LP codes with almost linear minimum distance, as well as an investigation of the performance of non-binary decoders for these codes. Specifically, we build on the work of~\cite{borah2024nonbinary} and extend the stacked hypergraph product construction presented there to a product of matrices over the $\F$-algebra $\F[x]/\langle x^\ell - 1\rangle$. This allows us to define a qudit lifted product construction in which the lift is taken over an $\F$-algebra rather than over the elements of $\F$ itself. The new construction contains the stacked hypergraph product of~\cite{borah2024nonbinary} as the special case $\ell = 1$, and it yields a valid qudit CSS code for every stacking order $1 \le s \le m$. Since the lift is taken over the cyclic group, all codes obtained in this way are quasi-cyclic. Unlike the merged lifted product considered in~\cite{borah2024nonbinary}, it does not reduce to a stacked hypergraph product. We then construct explicit families of such codes by taking one factor of the product to be the lift of a $q$-ary Tanner code on a cyclic lift of a Ramanujan graph, with local codes whose duals also have linear distance. For these families, we derive an exact expression for the number of logical qudits. We also prove a minimum distance bound $d \ge \gamma \ell$ that holds for any $b \in \F[x]$ with $b(0) \neq 0$, without requiring $\gcd(\mathrm{ord}(b), p) = 1$. For $b = x - 1$, this yields families with $k = \Theta(\log N)$ and $d = \Theta(N/\log N)$. These distance results apply to the fully stacked case $s = m$; bounding the minimum distance of partially stacked codes remains open. In addition, we complement the rate bounds of~\cite{borah2024nonbinary} with a lower bound on the minimum distance of the fully stacked hypergraph product, which was left open in that work.

On the decoding side, the codes in~\cite{borah2024nonbinary} were evaluated using a CSS variant of the belief propagation decoder of Lai and Kuo~\cite{lai2021log}. In this work, we present Monte Carlo simulation benchmarks for both the hypergraph product as well as the lifted product constructions. We also identify a key limitation of this decoder when applied to stacked codes: stacking introduces a large number of 4-cycles into the Tanner graph. This explains the saturation of the FER gains at larger field sizes observed in~\cite{borah2024nonbinary}. To overcome this limitation for fully stacked codes, we propose a merged syndrome decoder, which runs a standard non-binary BP decoder on the unstacked Tanner graph. We provide a direct comparison between the two decoders which demonstrates that the merged syndrome decoder improves the frame error rate (FER) by up to roughly an order of magnitude over the binary syndrome decoder. Finally, we provide an explanation for the qudit advantage through the cycle determinant of a 4-cycle. We prove that a syndrome-based symbol-flipping decoder oscillates on a 4-cycle if and only if its cycle determinant is zero. This is always the case for binary codes, but it occurs only with probability $1/(q-1)$ when the edge labels are random non-zero elements of $GF(q)$.

The rest of this paper is organized as follows. Sec.~\ref{sec:background} reviews qudit Pauli operators, the split and merge operators, the qudit stabilizer formalism, and trace-additive codes, following~\cite{borah2024nonbinary}. Sec.~\ref{sec:codes} revisits the stacked qudit hypergraph product construction and its encoding rate, and establishes a bound on the minimum distance of the fully stacked construction. Sec.~\ref{sec:qudit-lp} introduces the qudit lifted product construction over $\F$-algebras and establishes that it yields valid CSS codes. Sec.~\ref{sec:good-lp} gives an explicit construction of qudit lifted product codes from lifted Tanner codes on expander graphs, and analyzes their number of logical qudits and their minimum distance. Sec.~\ref{sec:decoder} describes the trace-additive BP decoder and the merged syndrome decoder for fully stacked codes. Sec.~\ref{sec:results} presents numerical benchmarks and the cycle-determinant analysis of decoder failures. Sec.~\ref{sec:conclusion} concludes the paper and discusses open problems.

\section{Background}
\label{sec:background}
In this section, we provide a review of the necessary background material for the subsequent sections in this paper. We begin with an introduction to the non-binary Pauli operators that shall serve as a basis set for the set of all errors that may occur in a single qudit system. Subsequently, we provide a review of the non-binary stabilizer framework for qudit error correction. Finally, we conclude with a collection of results from the theory of classical additive codes, which we shall later use for establishing bounds on the rates of non-binary QLDPC code families.

\subsection{Qudit Pauli Operators}
\label{sec:qudit-pauli}
For binary stabilizer codes, the set of Pauli operators, $\mathcal{P}_2 = \{I, X, Y, Z\}$, forms a basis set for the set of all errors that can occur in a single-qubit system. For a $q$-dimensional qudit, where $q$ is a prime power $(q = p^m)$, such a basis set is derived from the Galois field of order $q$, which we denote as $GF(q)$. In this representation, the Hilbert space of a single qudit of dimension $q = p^m$ consists of basis elements indexed by elements from $GF(q)$. Thus, an orthonormal basis for a single qudit Hilbert space is specified as,
\begin{equation}
    \mathcal{B} = \{|\eta\rangle: \eta \in GF(q)\}.
\end{equation}
Physically, these quantum states could refer to any set of orthonormal states in the physical qudit system. For example, our qudit could be a superconducting quantum system modeled as a quantum harmonic oscillator \cite{blais2021circuit}, and we could choose the states $\{|\eta\rangle\}$ to be distinct GKP qudit states \cite{schmidt2022quantum}. For this work, we shall make no assumptions about the physical states that represent the states $\{|\eta\rangle\}$. However, it should be noted that decoder performance may sometimes be significantly improved by using soft information obtained from these physical states \cite{raveendran2022finite}. Before proceeding to define the non-binary Pauli operators for qudit systems, we need to define the field trace function in a Galois field. This is done below.
\begin{definition}
\label{def:trace}
The field trace of an extension field $GF(p^m)$ over the base field $GF(p)$ is defined as follows.
\begin{equation}
    tr(\eta) = \eta + \eta^p + \eta^{p^2} + \dots + \eta^{p^{m-1}}.
\end{equation}
\end{definition}
Note that the field trace is a function from $GF(p^m)$ to $GF(p)$, i.e., $tr(\eta) \in GF(p)$ for any $\eta \in GF(p^m)$. Having defined qudit states and the field trace, we may then proceed to define the non-binary Pauli operators for a qudit of dimension $q$ as follows \cite{ketkar2006nonbinary}.
\begin{equation}
    X^a|x\rangle = |a + x\rangle, \qquad
    Z^b|x\rangle = e^{\frac{2 \pi i}{p}tr(bx)}|x\rangle,
    \label{eq:qudit-pauli}
\end{equation}
where $a, b, x \in GF(q)$. We note that all operations in the above equation are with respect to the field operations in $GF(q)$. The set of non-binary Pauli operators for a qudit of dimension $q$ is thus given by $\mathcal{P}_q = \{X^a Z^b: a, b \in GF(q)\}$. A qudit-Pauli operator $X^a Z^b \in \mathcal{P}_q$ is often represented by the pair $[a|b] \in GF(q)^2$. This notation is referred to as the symplectic notation \cite{nielsen2010quantum}. Similarly, an $N$-fold tensor product of Pauli operators acting on an $N$-qudit system has the following symplectic representation,
\begin{equation}
    \bigotimes_{i=1}^N X^{a_i} Z^{b_i} \mapsto [\; a_1 \; a_2 \;\dots \; a_N \;|\; b_1 \; b_2 \;\dots \; b_N\;].
\end{equation}

\begin{definition}
\label{def:pauli-group}
The single qudit Pauli group for a $q$-dimensional system, where $q = p^m$, is defined as follows.
\begin{equation}
    \mathcal{G}_{q} = \{e^{\frac{2 \pi i}{p} x} X^a Z^b: x \in GF(p); a, b \in GF(q)\}.
\end{equation}
The Pauli group for a system of $N$ qudits is defined as the $N$-fold tensor product of the single qudit Pauli group and is denoted by $\mathcal{G}_{q}^{\otimes N}$.
\end{definition}
An interesting observation in relation to Pauli operators in $GF(q)$ is that a single qudit of dimension $q = p^m$, for some prime $p$, can be visualized as a collection of $m$ sub-qudits, each of dimension $p$. To see this, we note that $GF(q)$ is isomorphic to a vector space of dimension $m$ over $GF(p)$ \cite{mceliece2012finite}. Then, there exists a basis $B_\alpha = \{\alpha_0, \alpha_1, \dots, \alpha_{m-1}\}$ of $GF(q)$ over $GF(p)$.
\begin{definition}
\label{def:dual-basis}
Let $q = p^m$ and $B_\alpha = \{\alpha_0, \alpha_1, \dots, \alpha_{m-1}\}$ be any basis of $GF(q)$ over $GF(p)$. Then, another basis set $B_\beta = \{\beta_0, \beta_1, \dots, \beta_{m-1}\}$ of $GF(q)$ over $GF(p)$ is called the dual basis of the primal basis $B_\alpha$ if
\begin{equation}
    tr(\alpha_i \beta_j) = \delta_{ij}; \quad 0 \leq i, j < m.
\end{equation}
\end{definition}
It is a well-known fact of Galois theory that the dual basis for any given primal basis always exists and is unique \cite{mceliece2012finite}. Let $\omega$ be a primitive element in $GF(q)$. For the purposes of this work, it will suffice to consider the primal basis $B_\alpha$ to be equal to the polynomial basis $\{1, \omega, \omega^2, \dots, \omega^{m-1}\}$ and the basis $B_\beta$ to be equal to its dual. In other words, unless specified otherwise, we will consider $\alpha_i = \omega^i$. The concept of primal and dual bases is an important tool that will be used consistently throughout this work. As an important application, consider an arbitrary element $\eta \in GF(q)$. This element can be expanded in either of the two bases as follows,
\begin{equation}
    \eta = \sum_{i=0}^{m-1} a_i \alpha_i = \sum_{i=0}^{m-1} b_i \beta_i,
\end{equation}
where each $a_i, b_i \in GF(p)$. Now, note from Def. \ref{def:dual-basis} that we must have $tr(\eta\beta_i) = a_i$ and $tr(\eta\alpha_i) = b_i$. Thus, the coefficients of any $\eta \in GF(q)$ with respect to any of the two bases can be extracted by taking a product with the corresponding dual basis elements and subsequently taking the field trace. Specifically, for any $\eta \in GF(q)$, we have that,
\begin{equation}
    \label{eq:basis-components}
    \eta = \sum_{i=0}^{m-1} tr(\eta\beta_i) \alpha_i = \sum_{i=0}^{m-1} tr(\eta\alpha_i) \beta_i.
\end{equation}

\begin{definition}
\label{def:split-op}
Let $q = p^m$ for some prime $p$. Let $B_\alpha = \{\alpha_0, \alpha_1, \dots, \alpha_{m-1}\}$ be a basis of $GF(q)$ over $GF(p)$ and let $B_\beta = \{\beta_0, \beta_1, \dots, \beta_{m-1}\}$ be its dual basis. The split operator with respect to $B_\alpha$ is the map $\psi_\alpha : GF(q) \to GF(p)^m$ defined as follows.
\begin{equation}
    \psi_\alpha(x) = \begin{bmatrix}
        tr(x\beta_0) & tr(x\beta_1) & \dots & tr(x\beta_{m-1})
    \end{bmatrix}.
\end{equation}
\end{definition}
In other words, the split operator $\psi_\alpha$ transforms any $\eta \in GF(q)$ into its coefficients over $GF(p)$ with respect to the basis $B_\alpha$. The split operator $\psi_\beta$ can also be defined similarly with respect to the dual basis. The merge operator defined next is the inverse of the split operator.
\begin{definition}
\label{def:merge-op}
Let $q = p^m$ for some prime $p$. Let $B_\alpha = \{\alpha_0, \alpha_1, \dots, \alpha_{m-1}\}$ be a basis of $GF(q)$ over $GF(p)$. The merge operator with respect to $B_\alpha$ is the map $\phi_\alpha : GF(p)^m \to GF(q)$ defined as follows.
\begin{equation}
\begin{split}
    & \;\phi_\alpha(\begin{bmatrix}
        a_0 & a_1 & \dots & a_{m-1}
    \end{bmatrix}) \\ 
    &= a_0 \alpha_0 + a_1 \alpha_1 + \dots + a_{m-1} \alpha_{m-1}.
\end{split}
\end{equation}
\end{definition}
With these definitions at hand, we can now express the elements $a, x \in GF(q)$ in Eq. \ref{eq:qudit-pauli} as a sum of the basis elements in $B_\alpha$. Similarly, we express $b \in GF(q)$ as a sum over the basis elements in $B_\beta$.
\begin{equation}
    a = \sum_{i=0}^{m-1} a_i \alpha_i; \qquad
    b = \sum_{i=0}^{m-1} b_i \beta_i; \qquad
    x = \sum_{i=0}^{m-1} x_i \alpha_i,
\end{equation}
where each $a_i, b_i, x_i \in GF(p)$. Then, plugging in the above expressions into Eq. \ref{eq:qudit-pauli}, we get,
\begin{equation}
\begin{split}
    X^a|x\rangle &= \left|\sum_{i=0}^{m-1} (a_i + x_i) \alpha_i \right\rangle, \\
    Z^b|x\rangle &= \prod_{i=0}^{m-1} e^{\frac{2 \pi i}{p} x_i b_i} \left|\sum_{i=0}^{m-1} x_i \alpha_i\right\rangle,
    \label{eq:dual-split}
\end{split}
\end{equation}
where we have used the duality of the bases $B_\alpha$ and $B_\beta$ in the second of the two equations above. Finally, we will make the following transformation to show the correspondence between $GF(q)$ and $GF(p)^m$.
\begin{equation}
    \psi : |x\rangle \mapsto |x_0\rangle \otimes |x_1\rangle \otimes \dots \otimes |x_{m-1}\rangle.
    \label{eq:subqudit-transform}
\end{equation}
Under this transformation, Eq. \ref{eq:dual-split} becomes,
\begin{equation}
\begin{split}
    \label{eq:gfp-gfq-equiv}
    \psi : X^a|x\rangle &\mapsto \bigotimes_{i=0}^{m-1} |a_i + x_i\rangle = \bigotimes_{i=0}^{m-1} X^{a_i}|x_i\rangle, \\
    \psi : \,Z^b|x\rangle &\mapsto \bigotimes_{i=0}^{m-1} e^{\frac{2 \pi i}{p} x_i b_i} \left|x_i\right\rangle = \bigotimes_{i=0}^{m-1} Z^{b_i}|x_i\rangle.
\end{split}
\end{equation}
Hence, we conclude from the above equation that the transformation defined in Eq. \ref{eq:subqudit-transform} preserves the action of both the $X$-type as well as the $Z$-type Pauli operators and so, we see that a single $q$-dimensional qudit is equivalent to $m$ distinct $p$-dimensional sub-qudits. Specifically, the transformation $\psi$ introduced above has the following effect on Pauli operators and qudit states.
\begin{equation}
\begin{split}
    \psi &: X^a Z^b \mapsto \bigotimes_{i=0}^{m-1} X^{tr(a\beta_i)} Z^{tr(b\alpha_i)} \\
    \psi &: |x\rangle \mapsto \bigotimes_{i=0}^{m-1} |tr(x\beta_i)\rangle.
\end{split}
\end{equation}
The transformation $\psi$ is therefore a group isomorphism between the Pauli groups $\mathcal{G}_q$ and $\mathcal{G}_p^{\otimes m}$. Under the symplectic notation, the isomorphism $\psi : \mathcal{G}_q \to \mathcal{G}_p^{\otimes m}$ may be written as follows.
\begin{equation}
    \label{eq:split}
    \psi : [\;a\;|\;b\;] \mapsto [\; \psi_\alpha(a) \;\;|\;\; \psi_\beta(b) \;],
\end{equation}
where $\psi_\alpha$ and $\psi_\beta$ are the split operators defined in Def. \ref{def:split-op}. The inverse of the above isomorphism is the isomorphism $\phi : \mathcal{G}_p^{\otimes m} \to \mathcal{G}_q$, which may be written in terms of the merge operators $\phi_\alpha$ and $\phi_\beta$ as follows.
\begin{equation}
\begin{split}
    \label{eq:merge}
    &\phi : [a_0 \;\; a_1 \;\; \dots \;\; a_{m-1} \;\;|\;\; b_0 \;\; b_1 \;\; \dots \;\; b_{m-1}] \\
    &\mapsto [\phi_\alpha(a_0, a_1, \dots, a_{m-1}) | \phi_\beta(b_0, b_1, \dots, b_{m-1})].
\end{split}
\end{equation}
In summary, we have proved the following result.
\begin{theorem}
\label{thm:equiv}
The group of Pauli operators $\mathcal{G}_q$ acting on a single qudit of dimension $q = p^m$ is isomorphic to the group of Pauli operators $\mathcal{G}_p^{\otimes m}$ acting on a collection of $m$ sub-qudits, each of dimension $p$. The split $(\psi)$ and merge $(\phi)$ transforms defined in Eqs. \ref{eq:split} and \ref{eq:merge} respectively, provide the necessary isomorphisms for transformations between these two groups.
\end{theorem}

\subsection{Qudit Stabilizer Codes}
\label{sec:qudit-stab}
Having defined the qudit Pauli operators and the qudit Pauli group for multilevel quantum systems, we are now in a position to define qudit stabilizer codes.
\begin{definition}
A qudit stabilizer group $\mathcal{S}$, on a set of $N$ qudits, is an Abelian subgroup of the $N$-fold Pauli group $\mathcal{G}_q^{\otimes N}$. Let $S_1, S_2, \dots, S_M$ be a set of generators for $\mathcal{S}$, so that $\mathcal{S} = \langle S_1, S_2, \dots, S_M \rangle$. The stabilizer code $\mathcal{C}$ defined by $\mathcal{S}$ may then be defined as the subspace of the $N$-qudit Hilbert space satisfying the condition
\begin{equation}
    S_i |\psi\rangle = |\psi\rangle \;\forall\; i = 1, 2, \dots, M.
\end{equation}
\end{definition}
The condition for commutativity is necessary because the generators of the stabilizer code are required to be simultaneously observable in order to make valid syndrome measurements \cite{nielsen2010quantum}. Commutativity for the qudit Pauli operators is described via the following theorem, the proof of which may be found in \cite{ketkar2006nonbinary}.
\begin{theorem}
\label{thm:commutativity}
Let $X^{x_1} Z^{z_1}$ and $X^{x_2} Z^{z_2}$ be two arbitrary single qudit Pauli operators. The commutativity of these two operators is described by the following relation.
\begin{equation}
    X^{x_1} Z^{z_1} X^{x_2} Z^{z_2} = e^{\left[\frac{2\pi i}{p} tr(z_1 x_2 - x_1 z_2)\right]} X^{x_2} Z^{z_2} X^{x_1} Z^{z_1}.
\end{equation}
In particular, $X^{x_1} Z^{z_1}$ and $X^{x_2} Z^{z_2}$ commute if and only if $tr(z_1 x_2 - x_1 z_2) = 0$. For two $N$-qudit Pauli operators with symplectic representations $[\mathbf{x}_1|\mathbf{z}_1]$ and $[\mathbf{x}_2|\mathbf{z}_2]$, the same statement holds with $tr(z_1 x_2 - x_1 z_2)$ replaced by $tr(\mathbf{z}_1 \mathbf{x}_2^T - \mathbf{x}_1 \mathbf{z}_2^T)$.
\end{theorem}
Note that the stabilizer group $\mathcal{S}$ is fully defined by the generators $\langle S_1, S_2, \dots, S_M\rangle$. We denote each $S_i$ as follows.
\begin{equation}
    S_i = S_{i1} \otimes S_{i2} \otimes \dots \otimes S_{iN},
\end{equation}
where each $S_{ij} \in \mathcal{G}_q$. Following the symplectic notation, we represent each $S_{ij}$ as $[x_{ij}|z_{ij}]$ such that $S_{ij} = X^{x_{ij}}Z^{z_{ij}}$ up to a phase of $e^{\frac{2\pi i x}{p}}$ for some $x \in GF(p)$. We define $A_x$ as the matrix containing $x_{ij}$ as its entries and $A_z$ as the matrix containing $z_{ij}$ as its entries. The code $\mathcal{C}$ is then fully defined by the symplectic matrix $S = \left[A_x|A_z\right]$. For valid quantum codes, the group $\mathcal{S}$ needs to be Abelian, which, in the symplectic notation, translates to the condition,
\begin{equation}
    tr(A_z A_x^T - A_x A_z^T) = 0.
\end{equation}
\begin{note*}
The $tr$ function used above should not be confused with the matrix trace. In the above equation, $tr$ refers to the Galois field trace defined in Def. \ref{def:trace}, applied element-wise to the matrix within the brackets. In the following, unless otherwise specified, we will always take $tr$ to denote the element-wise Galois field trace applied to its argument.
\end{note*}
\begin{definition}
A CSS code is a special class of stabilizer codes where each generator $S_{ij}$ contains only $X$ operators or only $Z$ operators. In this case, the symplectic matrix for the code may be written as,
\begin{equation}
    S = \left[\begin{array}{c | c}
        H_x & 0  \\
        0 & H_z
    \end{array}\right]; \;\; tr(H_x H_z^T) = 0.
    \label{eq:css}
\end{equation}
\end{definition}

\subsection{Trace Additive Codes}
\label{sec:background:additive}
\label{sec:additive-codes}
A linear code is an error-correcting code over $GF(q)$, for some $q = p^m$, that is closed under addition of codewords as well as under multiplication by arbitrary elements from $GF(q)$. In contrast, an additive code is one which is only closed under addition. Additive codes are important in the context of qudit codes because the rate and distance properties of qudit CSS codes defined by parity check matrices $H_x$ and $H_z$, are related to the rate and distance properties of the classical additive code defined by $H_x, H_z$, rather than the properties of the linear code defined by these two matrices. The difference between the additive and linear codes defined by a parity check matrix is clarified in the following definition.
\begin{definition}
    \label{def:par-additive}
    Let $H$ be an $r \times n$ parity check matrix with entries from $GF(q)$. Then the linear code $\mathcal{L}$ and the additive code $\mathcal{A}$ defined by the parity check matrix $H$ are given as follows.
    \begin{equation}
    \begin{split}
        \mathcal{L} &= \{x \in GF(q)^n : Hx^T = 0\} \\
        \mathcal{A} &= \{x \in GF(q)^n : tr(Hx^T) = 0\}.
    \end{split}
    \end{equation}
\end{definition}
Intuitively, the trace condition is a direct consequence of Thm. \ref{thm:commutativity}. Because the commutativity of two elements in the qudit Pauli group is defined via the Galois field trace, it is natural for the quantum code defined by matrices $H_x, H_z$ to be related to the additive code defined by these matrices rather than the linear code defined by these matrices \cite{nadkarni2021mathbb}. It is also possible to define additive codes via generator matrices rather than parity check matrices.
\begin{definition}
    \label{def:gen-additive}
    Let $G$ be a $k \times n$ generator matrix with entries from $GF(q)$. Then the linear code $\mathcal{L}$ and the additive code $\mathcal{A}$ defined by the generator matrix $G$ are given by the following.
    \begin{equation}
    \begin{split}
        \mathcal{L} &= \{mG : m \in GF(q)^k\} \\
        \mathcal{A} &= \{mG : m \in GF(p)^k\}.
    \end{split}
    \end{equation}
\end{definition}
Let $H_x$ and $H_z$ be non-binary matrices with entries from $GF(q)$ with dimensions $r_x \times n$ and $r_z \times n$ respectively. Let $CSS(H_x, H_z)$ be the qudit CSS code defined by $H_x, H_z$ and let $\mathcal{A}_x, \mathcal{A}_z$ be the respective (classical) additive codes defined by these parity check matrices. It is a well-known fact that the dual additive codes $\mathcal{A}_x^{\perp}, \mathcal{A}_z^{\perp}$ are generated additively by $H_x, H_z$ respectively. Specifically, we have,
\begin{equation}
\begin{split}
    \mathcal{A}_x &= \{v \in GF(q)^n : tr(H_x v^T) = 0\}, \\ 
    \mathcal{A}_z &= \{v \in GF(q)^n : tr(H_z v^T) = 0\}, \\
    \mathcal{A}_x^{\perp} &= \{m H_x : m \in GF(p) ^ {r_x}\}, \\
    \mathcal{A}_z^{\perp} &= \{m H_z : m \in GF(p) ^ {r_z}\}.
\end{split}
\end{equation}
From the above equations, it is easily observed that under the symplectic notation, $\mathcal{A}_x^{\perp}$ denotes the subgroup of $X$-type stabilizers in $CSS(H_x, H_z)$. Similarly, $\mathcal{A}_z^{\perp}$ denotes the subgroup of $Z$-type stabilizers. On the other hand, $\mathcal{A}_x$ is the set of all $n$-fold Pauli $Z$-type operators that commute with all the $X$-type stabilizers, and $\mathcal{A}_z$ is the set of all $n$-fold Pauli $X$-type operators that commute with all the $Z$-type stabilizers. The logical operators of the qudit CSS code are defined as the set of operators that commute with all the stabilizers but are themselves not within the stabilizer group. Thus, the $X$-type and $Z$-type logical operators of the CSS code are given by the following factor groups.
\begin{equation}
    \label{eq:factor-grp}
    L_x = \mathcal{A}_z / \mathcal{A}_x^{\perp}; \qquad L_z = \mathcal{A}_x / \mathcal{A}_z^{\perp}.
\end{equation}
It is well-known that the number of $X$-type logicals is equal to the number of $Z$-type logicals in a CSS code \cite{nielsen2010quantum}. In other words, $|L_x| = |L_z|$.
\begin{definition}
    Let $CSS(H_x, H_z)$ be an $n$-qudit CSS code over $GF(q)$ with $X$-type logicals $L_x$, and $Z$-type logicals $L_z$. Then, the encoding rate of the CSS code is defined as follows.
    \begin{equation}
        \mathcal{R} = \frac{1}{n} \log_q |L_x| = \frac{1}{n} \log_q |L_z|.
    \end{equation}
\end{definition}
\noindent Using Eq. \ref{eq:factor-grp}, this simplifies to,
\begin{equation}
\begin{split}
    \label{eq:rate-formula}
    \mathcal{R} &= \frac{1}{n} \left(\log_q |\mathcal{A}_x| - \log_q |\mathcal{A}_z^{\perp}|\right) \\ 
    &= \frac{1}{n} \left(\log_q |\mathcal{A}_z| - \log_q |\mathcal{A}_x^{\perp}|\right).
\end{split}
\end{equation}

\section{Qudit Hypergraph Product Codes}
\label{sec:codes}
We now proceed to the construction of qudit LDPC codes over finite fields of higher dimensions. We begin with an outline of the process involved in constructing qudit hypergraph product and qudit lifted product codes. Subsequently, we develop bounds on the encoding rate of the stacked hypergraph product construction, and state a lower bound on the minimum distance of its fully stacked version. The minimum distance of the lifted product construction is addressed in Sec.~\ref{sec:good-lp}.

\subsection{Qudit HGP Construction}
\label{sec:codes:hypergraph_product}

Consider a finite field $\mathbb{F}_q$, which is an order $m$ extension of the field $\mathbb{F}_p$, i.e., $q=p^m$. Let $B_{\alpha} = \{\alpha_0, \alpha_1, \ldots, \alpha_{m-1}\}$ be the primal basis of $\mathbb{F}_q$ as a vector space over $\mathbb{F}_p$ and  $B_{\beta} = \{\beta_0, \beta_1, \ldots, \beta_{m-1}\}$ be the trace-dual basis of $B_{\alpha}$.
Consider the stacking operators $\Omega_{\alpha,s}$ and $\Omega_{\beta,s}$ defined as follows:
\begin{eqnarray}
\label{eq:partial-stack-op}
\Omega_{\alpha,s} & = & \begin{bmatrix} \alpha_0 & \alpha_1 & \ldots & \alpha_{s-1}\end{bmatrix}^T \\
\Omega_{\beta,s} & = & \begin{bmatrix} \beta_0 & \beta_1 & \,\ldots\, & \beta_{s-1}\end{bmatrix}^T.
\end{eqnarray}
Let $H_1$ and $H_2$ denote the parity-check matrices of two classical codes over $\mathbb{F}_q$, with sizes $r_1 \times n_1$ and $r_2 \times n_2$ respectively. 
The order-$s$ stacked hypergraph product construction is given by the following matrices constituting the $X$-type and $Z$-type stabilizers:
\begin{eqnarray}
H_x & = & \Omega_{\alpha,s} \otimes H_{x0} \\
& = & \Omega_{\alpha,s} \otimes \begin{bmatrix}+H_1 \otimes I_{n_2} & +I_{r_1} \otimes H_2^T \end{bmatrix}, \\
H_z & = & \Omega_{\beta,s} \otimes H_{z0} \\
& = & \Omega_{\beta,s} \otimes \begin{bmatrix}+I_{n_1} \otimes H_2 & -H_1^T \otimes I_{r_2} \end{bmatrix},
\end{eqnarray}
where $H_{x0}, H_{z0}$ are the check matrices of the unstacked hypergraph product code.

\subsection{Encoding Rate}
\label{sec:params:rate}
From Eq. \ref{eq:rate-formula}, we note that the encoding rate for a qudit CSS code defined by check matrices $H_x, H_z$ is governed by the cardinalities of the additive codes $\mathcal{A}_x, \mathcal{A}_z, \mathcal{A}_x^{\perp}, \mathcal{A}_z^{\perp}$ defined by these matrices. Hence, we begin by developing an expression for the size of an additive code $\mathcal{A}$ and its dual as a function of the rank of the parity check matrix $H$ defining the code.
\begin{theorem}
    \label{thm:additive-size}
    Let $\mathcal{A}$ be an additive code defined by an $r \times n$ parity check matrix $H$ over $GF(q)$, where $q=p^m$. Then, the sizes of $\mathcal{A}$ and $\mathcal{A}^{\perp}$ are given by,
    \begin{equation}
        |\mathcal{A}| = p^{mn - \text{\normalfont rk}(\psi_\alpha(H))}; \qquad |\mathcal{A}^{\perp}| = p^{\text{\normalfont rk}(\psi_\alpha(H))},
    \end{equation}
    where we use \text{\normalfont rk} to denote the rank of a matrix and $B_\alpha$ is an arbitrary basis of $GF(q)$ over $GF(p)$.
\end{theorem}
\begin{proof}
    Let $B_\alpha, B_\beta$ be a pair of primal and dual basis sets. Applying the split operator to Def. \ref{def:par-additive}, we get the following.
    \begin{equation}
    \begin{split}
        \mathcal{A} &= \{x \in GF(q)^n : tr(Hx^T) = 0\} \\
        &= \{x \in GF(q)^n : \psi_\alpha(H) \psi_\beta(x)^T = 0\} \\
        &= \{\phi_\beta(y) \in GF(q)^n : \psi_\alpha(H) y^T = 0\},
    \end{split}
    \end{equation}
    where we have set $y = \psi_\beta(x)$ and used the fact that $\phi_\beta$ is the inverse of $\psi_\beta$ (see Def. \ref{def:split-op}). Let $\mathcal{L}$ be the linear code defined by the parity check matrix $\psi_\alpha(H)$. Then, the above equation tells us that $\mathcal{A} = \phi_\beta(\mathcal{L})$. Because $\phi_\beta$ is a bijective function (its inverse is $\psi_\beta$), this tells us that $|\mathcal{A}| = |\mathcal{L}|$. Now, $\psi_\alpha(H)$ is an $r \times mn$ parity check matrix over $GF(p)$. So the theory of linear codes tells us that,
    \begin{equation}
        \label{eq:additive-code-size}
        |\mathcal{A}| = |\mathcal{L}| = p^{mn - \text{\normalfont rk}(\psi_\alpha(H))}.
    \end{equation}
    Next, we consider $\mathcal{A}^{\perp}$ as defined in Def. \ref{def:gen-additive}.
    \begin{equation}
    \begin{split}
        \mathcal{A}^{\perp} &= \{zH : z \in GF(p)^r\} \\
        &= \{z\phi_\alpha(\psi_\alpha(H)) : z \in GF(p)^r\} \\
        &= \{\phi_\alpha(z\psi_\alpha(H)) : z \in GF(p)^r\},
    \end{split}
    \end{equation}
    where we have again used the fact that $\phi_\alpha$ is the inverse of $\psi_\alpha$ and we are able to move $z \in GF(p)^r$ inside the $\phi_\alpha$ function because $\phi_\alpha$ is a group homomorphism from the additive group of $GF(p)^m$ to the additive group of $GF(q)$. Now, note that $\mathcal{L}^{\perp} = \{z\psi_\alpha(H) : z \in GF(p)^r\}$ is a linear code over $GF(p)$. The above equation says that $\mathcal{A}^{\perp} = \phi_\alpha(\mathcal{L}^{\perp})$. Therefore, we can conclude that,
    \begin{equation}
        \label{eq:additive-dual-size}
        |\mathcal{A}^{\perp}| = |\mathcal{L}^{\perp}| = p^{\text{\normalfont rk}(\psi_\alpha(H))}.
    \end{equation}
    From Eq. \ref{eq:additive-code-size} and Eq. \ref{eq:additive-dual-size}, the theorem follows.
\end{proof}
\begin{remark*}
    Note that in the above theorem, we could easily have replaced the basis $B_\alpha$ with an arbitrary $B_\gamma$ and all arguments would still hold. Thus, we arrive at the important conclusion that $\text{\normalfont rk}(\psi_\alpha(H))$ is independent of the basis $B_\alpha$ under which we choose to split the matrix $H$.
\end{remark*}
\begin{corollary}
    \label{cor:rate-css}
    Let $H_x, H_z$ be the parity check matrices over $GF(q)$ defining a qudit CSS code $CSS(H_x, H_z)$, where $q = p^m$. Then, the rate of the code $CSS(H_x, H_z)$ is given by,
    \begin{equation}
        \mathcal{R} = 1 - \frac{\text{\normalfont rk}(\psi_\alpha(H_x)) + \text{\normalfont rk}(\psi_\alpha(H_z))}{mn}.
    \end{equation}
\end{corollary}
\begin{proof}
    Direct consequence of Thm. \ref{thm:additive-size} and Eq. \ref{eq:rate-formula}.
\end{proof}
Let $H_1, H_2$ be two parity check matrices over $GF(q)$ of dimensions $r_1 \times n_1$ and $r_2 \times n_2$ respectively. Let us denote by $k_1, k_2, k_1^T, k_2^T$, the dimensions of the linear codes defined by $H_1, H_2, H_1^T, H_2^T$ respectively. For the case of $q=2$, it was shown in \cite{tillich2013quantum}, that the encoding rate $\mathcal{R}_0$ of the qubit hypergraph product code is,
\begin{equation}
    \label{eq:base-rate}
    \mathcal{R}_0 = \frac{k_1 k_2 + k_1^T k_2^T}{n_1 n_2 + r_1 r_2}.
\end{equation}
The same formula holds for $q=p$ with $p$ an arbitrary prime, since the proof in \cite{tillich2013quantum} uses only linear algebra over the field. However, for the case of $q=p^m$, where $m \geq 2$, the above equation no longer holds true. This is because the encoding rate of a qudit code is defined by the sizes of the additive codes $\mathcal{A}_x$ and $\mathcal{A}_z$ defined by $H_x, H_z$ rather than the linear codes defined by these matrices (see Eq. \ref{eq:rate-formula}). The following theorem provides bounds on the encoding rate $\mathcal{R}$ of the stacked hypergraph product code for arbitrary $q$.
\begin{theorem}
    \label{thm:shp-rate}
    Let $H_1, H_2$ be two parity check matrices over $GF(q)$, where $q=p^m$ for some prime $p$. Then, the encoding rate $\mathcal{R}$ of the order-$s$ stacked hypergraph product code obtained from these matrices is bounded as below.
    \begin{equation}
        \mathcal{R}_0 \leq \mathcal{R} \leq 1 - \frac{s}{m}(1 - \mathcal{R}_0),
    \end{equation}
    where $\mathcal{R}_0$ is given by Eq. \ref{eq:base-rate}.
\end{theorem}
\begin{proof}
    The order-$s$ stacked hypergraph product code obtained from $H_1, H_2$ is given by the check matrices $H_x = \Omega_{\alpha,s} \otimes H_{x0}$ and $H_z = \Omega_{\beta,s} \otimes H_{z0}$, where $\Omega_{\alpha,s}, \Omega_{\beta,s}$ are given by Eq. \ref{eq:partial-stack-op} and $H_{x0}, H_{z0}$ are given by,
    \begin{equation}
    \begin{split}
        \label{eq:nb-hpc}
        H_{x0} &= \left[+H_1 \otimes I_{n_2} \quad +I_{r_1} \otimes H_2^T \right], \\
        H_{z0} &= \left[+I_{n_1} \otimes H_2 \quad -H_1^T \otimes I_{r_2} \right].
    \end{split}
    \end{equation}
    Let us define $\chi$ as the set of linearly independent rows in $H_{x0}$.
    \begin{equation}
        \label{eq:chi}
        \chi = \{x_1, x_2, \dots, x_{\text{\normalfont rk} (H_{x0})}\},
    \end{equation}
    where each $x_i$ is a row vector with entries from $GF(q)$. Then, we consider the following set of row vectors in the split version of $H_x$, given by $\psi_\alpha(H_x) = \psi_\alpha(\Omega_{\alpha,s} \otimes H_{x0})$.
    \begin{equation}
        \label{eq:chi-alpha}
        \chi_{\alpha, s} = \{\psi_\alpha(\alpha_i x_j): 0 \leq i \leq s-1, 1 \leq j \leq \text{\normalfont rk} (H_{x0})\}.
    \end{equation}
    It is easy to see that the above is a set of linearly independent vectors in $GF(p)^{mn}$. Thus, $\psi_\alpha(H_x)$ contains at least the elements in $\chi_{\alpha,s}$ as linearly independent rows. Because the set $\chi_{\alpha,s}$ has $s\cdot\text{\normalfont rk} (H_{x0})$ elements, we may infer,
    \begin{equation}
        \label{eq:rank-bound-x}
        s\cdot\text{\normalfont rk} (H_{x0}) \leq \text{\normalfont rk} (\psi_\alpha(H_{x})).
    \end{equation}
    A similar argument for $H_z$, along with the observation that $\text{\normalfont rk} (\psi_\alpha(H_{z}))$ does not depend on the basis $B_\alpha$, implies that,
    \begin{equation}
        \label{eq:rank-bound-z}
        s\cdot\text{\normalfont rk} (H_{z0}) \leq \text{\normalfont rk} (\psi_\alpha(H_{z})).
    \end{equation}
    Combining the two inequalities above with Cor. \ref{cor:rate-css} yields,
    \begin{equation}
        \label{eq:upper-bound-tmp}
        \mathcal{R} \leq 1 - \frac{s}{mn} (\text{\normalfont rk} (H_{x0}) + \text{\normalfont rk} (H_{z0})),
    \end{equation}
    where $n = n_1 n_2 + r_1 r_2$ is the number of physical qudits required by the stacked hypergraph product code. Let $k_{x0}, k_{z0}$ be the dimensions of the linear codes defined by $H_{x0}, H_{z0}$ respectively. Then, it was proved in Proposition 14 of \cite{tillich2013quantum} that,
    \begin{equation}
        k_{x0} + k_{z0} - n = k_1 k_2 + k_1^T k_2^T = n\mathcal{R}_0.
    \end{equation}
    From the theory of linear codes, we know that $k_{x0} = n - \text{\normalfont rk} (H_{x0})$ and $k_{z0} = n - \text{\normalfont rk} (H_{z0})$. Thus, the above equation reduces to,
    \begin{equation}
    \begin{split}
        & n - \text{\normalfont rk} (H_{x0}) - \text{\normalfont rk} (H_{z0}) = n\mathcal{R}_0 \\ \implies & \text{\normalfont rk} (H_{x0}) + \text{\normalfont rk} (H_{z0}) = n(1 - \mathcal{R}_0).
    \end{split}
    \end{equation}
    Applying the above equation to Eq. \ref{eq:upper-bound-tmp}, we obtain the upper bound as follows.
    \begin{equation}
        \label{eq:upper-bound}
        \mathcal{R} \leq 1 - \frac{s}{m} (1 - \mathcal{R}_0).
    \end{equation}
    To prove the lower bound, we begin with the observation that if $\mathcal{A}$ and $\mathcal{L}$ are respectively the additive and linear codes defined by a parity check matrix $H$, then we must have that $\mathcal{L} \subseteq \mathcal{A}$ and $\mathcal{A}^\perp \subseteq \mathcal{L}^\perp$. This is easily seen as a direct consequence of Def. \ref{def:par-additive} and Def. \ref{def:gen-additive}. Applying this to Eq. \ref{eq:rate-formula}, we get,
    \begin{equation}
    \begin{split}
        \mathcal{R} &= \frac{1}{n} \left(\log_q |\mathcal{A}_x| - \log_q |\mathcal{A}_z^{\perp}|\right) \\
        &\geq \frac{1}{n} \left(\log_q |\mathcal{L}_x| - \log_q |\mathcal{L}_z^{\perp}|\right).
    \end{split}
    \end{equation}
    In the above, $\mathcal{L}_x$ and $\mathcal{L}_z$ are the linear codes defined by $H_x, H_z$ respectively. Now, we note that these linear codes are exactly the same as the codes $\mathcal{L}_{x0}$ and $\mathcal{L}_{z0}$ defined respectively by $H_{x0}, H_{z0}$. This is simply a consequence of the fact that multiplying rows of $H_{x0}$ or $H_{z0}$ by basis elements $\alpha_i$ or $\beta_j$ does not generate any new linearly independent rows. Thus, we get,
    \begin{equation}
        \mathcal{R} \geq \frac{1}{n} \left(\log_q |\mathcal{L}_{x0}| - \log_q |\mathcal{L}_{z0}^{\perp}|\right).
    \end{equation}
    But then, we note that $\log_q |\mathcal{L}_{z0}^{\perp}| = n - k_{z0}$ and $\log_q |\mathcal{L}_{x0}| = k_{x0}$ and so, from Proposition 14 of \cite{tillich2013quantum}, the right side of the above inequality is simply $\mathcal{R}_0$. Thus, we get the lower bound as $\mathcal{R} \geq \mathcal{R}_0$, which concludes the proof of the theorem.
\end{proof}
\begin{corollary}
    \label{cor:full-stacked-rate}
    Let $H_1, H_2$ be two parity check matrices over $GF(q)$, where $q=p^m$ for some prime $p$. Then, the encoding rate $\mathcal{R}$ of the order-$m$ stacked hypergraph product code obtained from these matrices is given by $\mathcal{R} = \mathcal{R}_0$ as given by Eq. \ref{eq:base-rate}.
\end{corollary}
\begin{proof}
    Immediate consequence of Thm. \ref{thm:shp-rate} with $s=m$.
\end{proof}
It was shown in \cite{nadkarni2021mathbb} that stacking a parity check matrix $H$ to order $m$ yields an additive code defined by $H$ that is identical to the linear code defined by $H$. As such, the above corollary comes as no surprise and is a consequence of the fact that order-$m$ stacking simply serves to linearize the code defined by the parity check matrices $H_{x0}$ and $H_{z0}$. 

An important special case of Thm. \ref{thm:shp-rate} is the case when $H_1, H_2$ are full rank parity check matrices over $GF(q)$. Then, it is easy to see that $H_{x0}, H_{z0}$ are also full-rank (again, see Proposition 14 of \cite{tillich2013quantum}). In that case, the set $\chi$ as defined in Eq. \ref{eq:chi} consists of all the rows of $H_{x0}$ and the set $\chi_{\alpha,s}$ as defined in Eq. \ref{eq:chi-alpha} contains all rows of $\psi_\alpha(H_x)$. Consequently, the inequalities in Eq. \ref{eq:rank-bound-x} and Eq. \ref{eq:rank-bound-z} are saturated. Following through the above proof, it is easy to see that the upper bound in Eq. \ref{eq:upper-bound} is also saturated in this case. So, we arrive at the following theorem.
\begin{theorem}
    \label{thm:shp-frk-rate}
    Let $H_1, H_2$ be two full-rank parity check matrices over $GF(q)$, where $q=p^m$ for some prime $p$. Then, the encoding rate $\mathcal{R}$ of the order-$s$ stacked hypergraph product code obtained from these matrices is given by
    \begin{equation}
        \mathcal{R} = 1 - \frac{s}{m}(1 - \mathcal{R}_0),
    \end{equation}
    where $\mathcal{R}_0$ is given by Eq. \ref{eq:base-rate}.
\end{theorem}

\begin{theorem}
\label{thm:hgp-distance}
Let $d_1, d_2, d_1^T, d_2^T$ denote the minimum distances of the linear codes defined by $H_1, H_2, H_1^T, H_2^T$ respectively (with the convention that the distance of the trivial code is $\infty$). Then the minimum distance of the fully stacked ($s = m$) hypergraph product construction
satisfies $d \ge \min(d_1, d_2, d_1^T, d_2^T)$.
\end{theorem}

\begin{proof}
By Cor.~\ref{cor:full-stacked-rate} and \cite{nadkarni2021mathbb}, the fully stacked code is the $\mathbb{F}_q$-linear CSS code $CSS(H_{x0}, H_{z0})$, so its $Z$-distance is the minimum weight of a vector in $\ker H_{x0}$ that does not lie in the $GF(q)$-row space of $H_{z0}$, and similarly for the $X$-distance. The bound $d_z \ge \min(d_1, d_2^T)$ and $d_x \ge \min(d_2, d_1^T)$ on these quantities is the hypergraph product distance bound of \cite{tillich2013quantum}; its proof uses only that any $d-1$ columns of a parity-check matrix of a code of distance $d$ are linearly independent, together with linear algebra over the field, and therefore carries over from $GF(2)$ to $GF(q)$ unchanged.
\end{proof}


\section{Qudit Lifted Product Codes}
\label{sec:qudit-lp}
For non-binary lifted product codes, we construct the codes over group algebras.
Let $G$ be a finite group. The elements of the group algebra
$\F G$ are formal sums $\sum_{g\in G}\alpha_g g$ with $\alpha_g\in \F$. Addition is
coordinatewise, $\sum_g\alpha_g g+\sum_g\beta_g g=\sum_g(\alpha_g+\beta_g)g$, and
multiplication is given by
\[
  \Bigl(\sum_{g'}\alpha_{g'}g'\Bigr)\Bigl(\sum_{g''}\beta_{g''}g''\Bigr)
  =\sum_{g}\Bigl(\sum_{g',g'':\,g'g''=g}\alpha_{g'}\beta_{g''}\Bigr)g .
\]
After fixing an order on $G$, an element $\alpha\in \F G$ is identified with its
coefficient vector $\mathbb{b}(\alpha)\in \F^{\ell}$, $\ell=|G|$, regarded as a column
vector. Left multiplication by $\alpha$ is an $\F$-linear map on $\F G$, represented
in this basis by an $\ell\times\ell$ matrix $\rho(\alpha)$ over $\F$. The map $\rho$
is the left regular representation of $\F G$ and satisfies
\begin{itemize}
  \item $\mathbb{b}(\alpha\beta)=\rho(\alpha)\,\mathbb{b}(\beta)$,
  \item $\rho(\alpha+\beta)=\rho(\alpha)+\rho(\beta)$,
  \item $\rho(\alpha\beta)=\rho(\alpha)\rho(\beta)$.
\end{itemize}
In this work we only use the cyclic group $G=\mathbb{Z}_\ell$, for which $\F G$ is the
ring of polynomials modulo $x^\ell-1$ and $\rho(\alpha)$ is a circulant matrix.

\begin{definition}
\label{def:ring}
Let $\ell$ be a positive integer. We write
$R_\ell=\F[x]/\langle x^\ell-1\rangle$ for the ring of polynomials over $\F$ modulo
$x^\ell-1$. The \emph{antipode} of $\alpha=\sum_k\alpha_kx^k\in R_\ell$ is
$\bar\alpha=\sum_k\alpha_kx^{-k}$, and for a matrix $H$ over $R_\ell$ we write $\bar H$
for the entrywise antipode and $H^*=\bar H^{T}$ for the conjugate transpose.
We denote by $C_\ell\subset R_\ell$ the set of monomials of $R_\ell$,
\begin{equation}
  C_\ell=\{\,ax^k\in R_\ell : a\in \F,\ k\in\mathbb{Z},\ 0\le k<\ell\,\}.
\end{equation}
$C_\ell$ is closed under multiplication but not under addition; every element of
$R_\ell$ is a sum of at most $\ell$ elements of $C_\ell$.
\end{definition}

\begin{definition}
\label{def:lift}
Let $M_{\ell\times\ell}$ be the set of $\ell\times\ell$ matrices over $\F$ and let
$P\in M_{\ell\times\ell}$ be the cyclic shift matrix
\begin{equation}
  P=\begin{bmatrix}
      0 & 0 & \cdots & 0 & 1\\
      1 & 0 & \cdots & 0 & 0\\
      0 & 1 & \cdots & 0 & 0\\
      \vdots & \vdots & \ddots & \vdots & \vdots\\
      0 & 0 & \cdots & 1 & 0
    \end{bmatrix},
\end{equation}
obtained by applying a circular left shift to the rows of the identity matrix.
We define the lift operators $\rho,\rho^*:R_\ell\to M_{\ell\times\ell}$ on monomials by
\begin{equation}
  \rho(ax^k)=aP^k,\qquad \rho^*(ax^k)=a(P^T)^k=aP^{-k},
\end{equation}
and extend both to all of $R_\ell$ by $\F$-linearity. The operator $\rho$ coincides
with the left regular representation above, so it is a ring homomorphism, and
$\rho(\alpha)$ is the circulant matrix whose first column is $\mathbb{b}(\alpha)$.
The two lifts are related by
\begin{equation}\label{eq:rhostar}
  \rho^*(\alpha)=\rho(\bar\alpha)=\rho(\alpha)^T\qquad(\alpha\in R_\ell).
\end{equation}
Both operators are applied to a matrix over $R_\ell$ entrywise; if $H$ is
$r\times n$ over $R_\ell$ then $\rho(H),\rho^*(H)$ are $r\ell\times n\ell$ over $\F$.
By \eqref{eq:rhostar}, for any matrix $N$ over $R_\ell$,
\begin{equation}\label{eq:rhostarT}
  \rho^*(N)=\rho(\bar N),\qquad \rho^*(N)^T=\rho(N^T).
\end{equation}
\end{definition}

\begin{definition}
\label{def:lp}
Let $H_1$ and $H_2$ be matrices of sizes $r_1\times n_1$ and $r_2\times n_2$ with
entries from $R_\ell$. Let $q=p^m$ be the number of qudit levels, where $p$ is a prime
and $m$ is a positive integer, and let $1\le s\le m$ be an integer. The
\emph{order-$s$ qudit lifted product code} with respect to $H_1,H_2$ is the qudit CSS
code on $N=(n_1n_2+r_1r_2)\ell$ qudits with $X$- and $Z$-type matrices
\begin{equation}\label{eq:LPdef}
\begin{aligned}
  H_x &= \Omega_{\alpha,s}\otimes\rho\bigl(\bigl[\,+H_1\otimes I_{n_2}\quad +I_{r_1}\otimes H_2^T\,\bigr]\bigr),\\
  H_z &= \Omega_{\beta,s}\otimes\rho^*\bigl(\bigl[\,+I_{n_1}\otimes H_2\quad -H_1^T\otimes I_{r_2}\,\bigr]\bigr),
\end{aligned}
\end{equation}
where $\Omega_{\alpha,s},\Omega_{\beta,s}$ are the partial stacking operators of
Eq.~\eqref{eq:partial-stack-op} and $\rho,\rho^*$ are the lift operators of Definition~\ref{def:lift}, applied
entrywise. All products inside the brackets are carried out over $R_\ell$, with
$H^T$ denoting the plain transpose over $R_\ell$.
\end{definition}

\begin{proposition}
\label{prop:lp-css}
The matrices in \eqref{eq:LPdef} satisfy $tr(H_xH_z^T)=0$, so Definition~\ref{def:lp} is a valid
qudit CSS code for every $1\le s\le m$.
\end{proposition}

\begin{proof}
Let $M=[\,H_1\otimes I_{n_2}\ \ I_{r_1}\otimes H_2^T\,]$ and
$N=[\,I_{n_1}\otimes H_2\ \ -H_1^T\otimes I_{r_2}\,]$. By \eqref{eq:rhostarT} and the
fact that $\rho$ is a ring homomorphism on matrices over the commutative ring $R_\ell$,
\begin{eqnarray*}
  && \rho(M)\,\rho^*(N)^T=\rho(M)\,\rho(N^T)=\rho(MN^T),
  \\
 && MN^T=H_1\otimes H_2^T-H_1\otimes H_2^T=0 .
\end{eqnarray*}
Hence $H_xH_z^T=\Omega_{\alpha,s}\Omega_{\beta,s}^T\otimes\rho(MN^T)=0$, and in
particular $tr(H_xH_z^T)=0$.
\end{proof}

We note that the codes of Definition~\ref{def:lp} are quasi-cyclic. For every $\alpha\in R_\ell$, both $\rho(\alpha)$ and $\rho^*(\alpha)=\rho(\bar\alpha)$ are $\ell\times\ell$ circulant matrices (Definition~\ref{def:lift}). Hence the lifts of the bracketed matrices in \eqref{eq:LPdef} are block matrices whose blocks are $\ell\times\ell$ circulants. The stacking operators $\Omega_{\alpha,s}$ and $\Omega_{\beta,s}$ only replicate these block matrices and multiply each replica by a scalar $\alpha_i$ or $\beta_i\in\F$, which preserves the circulant property. Consequently, for every stacking order $1\le s\le m$, the matrices $H_x$ and $H_z$ consist entirely of $\ell\times\ell$ circulant blocks. Since a circulant block commutes with the cyclic shift $P$, the order-$s$ qudit lifted product code is invariant under simultaneous cyclic shifts of the coordinates within each block of $\ell$ qudits, i.e., it is quasi-cyclic with circulant size $\ell$. For $s<m$ the codes $\mathcal{A}_x,\mathcal{A}_z$ are additive rather than $\F$-linear, but the block-circulant structure of the parity-check matrices, which is what hardware decoder architectures exploit, is the same for every $s$.

\begin{remark}
\label{rem:lp-signs}
By \eqref{eq:rhostarT}, $H_z$ in \eqref{eq:LPdef} is the lift
$\Omega_{\beta,s}\otimes\rho\bigl([\,I_{n_1}\otimes\bar H_2\ \ -H_1^*\otimes I_{r_2}\,]\bigr)$:
the antipode enters $H_z$ through $\rho^*$. In the special case $H_1=A$,
$H_2=b\in R_\ell$ (a $1\times1$ matrix) used in Sec.~\ref{sec:good-lp}, Definition~\ref{def:lp} reduces to
\[
  H_x=\Omega_{\alpha,s}\otimes\rho\bigl([\,A\ \ bI_{r_1}\,]\bigr),\qquad
  H_z=\Omega_{\beta,s}\otimes\rho\bigl([\,\bar b\,I_{n_1}\ \ -A^*\,]\bigr),
\]
which for $s=m$ and $q=2$ is the code $\mathrm{LP}(A,b)$ of \cite{panteleev2021quantum}. Over $R_\ell$ a $Z$-type
vector $[u,v]$ thus lies in $\ker H_x$ iff $Au=-bv$, and the $Z$-type stabilizers are the
vectors $[bh,\,-Ah]$, $h\in R_\ell^{n_1}$. In Sec.~\ref{sec:good-lp} we use the sign convention
of \cite{panteleev2021quantum}, $Au=bv$ and $[bh,\,Ah]$, obtained from the above by $v\mapsto-v$; this does not
change any weight.
\end{remark}

\section{Qudit LP Codes with Almost Linear Minimum Distance}
\label{sec:good-lp}
In this section, we describe the construction of a qudit lifted product code
whose number of logical qudits grows as $\log N$ and whose minimum distance grows
as $\frac{N}{\log N}$, where $N$ is the block length. Throughout this section, the second factor of the lifted product is a single polynomial, which we
denote by $b\in R_\ell$; the symbol $g$ is reserved for $(x^\ell-1)/b$ in
Sec.~\ref{sec:lp-distance}. All codes in this section are fully stacked, i.e.\ the stacking order $s$ equals the extension degree of $\F$ over $\mathbb{F}_p$. Since the extension degree plays no further role here, in this section the symbol $m$ is reused for the number of rows of a matrix over $R_\ell$.

\begin{definition}
\label{def:tanner}
Let $\mathcal{G}=(V,E)$ be a $w$-regular graph on $2n_0$ vertices, so that
$|E|=wn_0$, and let $\mathcal{C}_0\subseteq\mathbb{F}_q^{w}$ be a linear code of
length $w$. For every $v\in V$ let $E(v)$ denote the set of $w$ edges incident to
$v$, and fix once and for all a bijection $E(v)\to\{1,\dots,w\}$. The
\emph{Tanner code} $\mathcal{T}(\mathcal{G},\mathcal{C}_0)\subseteq\mathbb{F}_q^{E}$
is the code of length $|E|=wn_0$ obtained by indexing coordinates by the edges
of $\mathcal{G}$ and imposing
\begin{equation}
  \forall\,\mathbf{c}\in\mathcal{T}(\mathcal{G},\mathcal{C}_0),\ \forall\,v\in V:\qquad
  \mathbf{c}|_{E(v)}\in\mathcal{C}_0,
\end{equation}
where $\mathbf{c}|_{E(v)}\in\mathbb{F}_q^{w}$ is read off through the fixed
bijection. If $H_0\in\mathbb{F}_q^{r\times w}$ is a parity-check matrix of
$\mathcal{C}_0$ whose rows form a basis of $\mathcal{C}_0^\perp$, the
\emph{canonical parity-check matrix} of $\mathcal{T}(\mathcal{G},\mathcal{C}_0)$
is the $2n_0r\times wn_0$ matrix $H$ whose rows are indexed by pairs
$(v,j)$, $v\in V$, $1\le j\le r$, with row $(v,j)$ equal to row $j$ of $H_0$
placed on the coordinates $E(v)$ and zero elsewhere.
\end{definition}

The construction is described in the following steps:
\begin{enumerate}
  \item We start with the construction in Definition~\ref{def:lp} and take $H_2$ to be the
        $1\times1$ matrix $b \in R_\ell$, such that $b | (x^{\ell}-1)$.
  \item We describe the construction of $H_1$ at high level first and then see
        each of the steps in detail.
  \begin{itemize}
    \item \textbf{Step 1:} The first ingredient is a code $\mathcal{C}_0$ over
          $\mathbb{F}_q$ of length $w$ such that both $\mathcal{C}_0$ and
          $\mathcal{C}_0^\perp$ have large minimum distance. Such codes are
          obtained by a random coding argument (Lemma~\ref{lem:gv}).
    \item \textbf{Step 2:} $\mathcal{C}_0$ is used as the local code of a Tanner
          code on an expander graph (Definition~\ref{def:tanner}). The canonical parity-check
          matrix of this Tanner code and its transpose are both expanding
          (Lemma~\ref{lem:expansion}).
    \item \textbf{Step 3:} The Tanner code is lifted through an $\ell$-cyclic
          group lift of the graph (Lemma~\ref{lem:lift}). The canonical parity-check matrix
          of the lifted Tanner code is $\mathbb{Z}_\ell$-equivariant, hence of the
          form $\rho(H_1)$ for a matrix $H_1$ over $R_\ell$, and $H_1$ is read off
          from $\rho(H_1)$ (Theorem~\ref{thm:h1}).
  \end{itemize}
\end{enumerate}

The following lemma shows that for sufficiently large $w$ there exist codes over
$\mathbb{F}_q$ such that both the code and its dual have a good relative distance.
Here $h_q(\delta)=\delta\log_q(q-1)-\delta\log_q\delta-(1-\delta)\log_q(1-\delta)$
is the $q$-ary entropy function.

\begin{lemma}
\label{lem:gv}
Let $\delta\in(0,1-q^{-1})$ and $\epsilon\in(0,\tfrac23)$ satisfy
\begin{equation}\label{eq:GVcond}
  \max\Bigl(\frac{\epsilon}{2},\ 1-\frac{\epsilon}{2}\Bigr)+\frac{\epsilon}{4}
  \ <\ 1-h_q(\delta).
\end{equation}
Then for all sufficiently large $w$ there exists a linear code
$\mathcal{C}_0\subseteq\mathbb{F}_q^{w}$ of dimension $w-r$ with
$\bigl|\frac{r}{w}-\frac{\epsilon}{2}\bigr|\le\frac{\epsilon}{4}$ such that
\begin{equation}
  \frac{d(\mathcal{C}_0)}{w}\ \ge\ \delta,\qquad
  \frac{d(\mathcal{C}_0^\perp)}{w}\ \ge\ \delta .
\end{equation}
\end{lemma}

\begin{proof}
Following \cite{barg2002random}, consider the ensemble in which an $r\times w$ parity-check matrix
$H_0$ is drawn uniformly at random. With probability $1-o(1)$ as $w\to\infty$ the
matrix $H_0$ has full rank $r$, so $\mathcal{C}_0=\ker H_0$ has rate
$R=1-\frac{r}{w}$ and $\mathcal{C}_0^\perp$, whose generator matrix is $H_0$, has
rate $1-R=\frac{r}{w}$. The result of \cite{barg2002random} states that a code from this ensemble
has relative distance at least $\delta$ with high probability whenever its rate is
below the Gilbert--Varshamov bound $1-h_q(\delta)$; it applies equally to the dual
code, since a uniformly random parity-check matrix of $\mathcal{C}_0$ is a uniformly
random generator matrix of $\mathcal{C}_0^\perp$. By the choice of $r$,
$R\le 1-\frac{\epsilon}{2}+\frac{\epsilon}{4}$ and
$1-R\le\frac{\epsilon}{2}+\frac{\epsilon}{4}$, so \eqref{eq:GVcond} places both rates
below $1-h_q(\delta)$. A union bound over the three events (full rank, distance of
$\mathcal{C}_0$, distance of $\mathcal{C}_0^\perp$) shows that all hold
simultaneously with high probability, which proves the lemma.
\end{proof}

\noindent {\bf Expanding $q$-ary matrices: }
A matrix $A\in\mathbb{F}_q^{M\times N}$ is said to be \emph{$(\alpha,\beta)$-expanding}
if for all $x\in\mathbb{F}_q^{N}$ with $|x|\le\alpha N$ it holds that $|Ax|\ge\beta|x|$.
If $A$ is $(\alpha,\beta)$-expanding then the code with parity-check matrix $A$ has
minimum distance greater than $\alpha N$. When this notion is applied to a lifted
matrix $\rho(A)$ with $A\in M_{m\times n}(R_\ell)$, the number of columns is
$N=n\ell$, so the condition reads $|x|\le\alpha\ell n$.

\begin{lemma}
\label{lem:expansion}
Let $\mathcal{C}_0\subseteq\mathbb{F}_q^{w}$ satisfy $\frac{d(\mathcal{C}_0)}{w}\ge\delta$
and $\frac{d(\mathcal{C}_0^\perp)}{w}\ge\delta$, let $\mathcal{G}$ be a
$(2n_0,w,\lambda)$ spectral expander with $\lambda<\delta w$, and let $H$ be the
canonical parity-check matrix of $\mathcal{T}(\mathcal{G},\mathcal{C}_0)$ of
Definition~\ref{def:tanner}. Then $H$ and $H^T$ are both $(\alpha,\beta)$-expanding for every
$\alpha<\frac{\delta}{w}\bigl(1-\frac{\lambda}{\delta w}\bigr)$, with
$\beta=\frac{1}{\delta w}\bigl(\delta-\alpha w-\frac{\lambda}{w}\bigr)>0$.
\end{lemma}

\begin{proof}
The proof is the same as that of Lemma~3 in \cite{panteleev2021quantum} with $\mathbb{F}_2$ replaced by  $\mathbb{F}_q$.
\end{proof}

\noindent {\bf $\ell$-cyclic group lift of a graph:}
Given a graph $\mathcal{G}=(V,E)$, an $\ell$-cyclic group lift of $\mathcal{G}$ is
obtained by replacing each vertex $v$ by the $\ell$ vertices $(v,i)$, $i\in\mathbb{Z}_\ell$,
and each edge by $\ell$ edges between the two sets of $\ell$ vertices, which form a
perfect matching defined by the action of an element of $\mathbb{Z}_\ell$. Explicitly,
orient every edge $e\in E$ as $e=(u\to v)$ and assign to it a shift $s(e)\in\mathbb{Z}_\ell$;
the lifted graph $\mathcal{G}^{(\ell)}$ has vertex set $V\times\mathbb{Z}_\ell$ and, for
every edge $e=(u\to v)$ and every $i\in\mathbb{Z}_\ell$, an edge $(e,i)$ joining $(u,i)$
to $(v,i+s(e))$. Traversing $e$ in the opposite direction uses the shift $-s(e)$.
The group $\mathbb{Z}_\ell$ acts freely on $\mathcal{G}^{(\ell)}$ by
$(v,i)\mapsto(v,i+1)$, $(e,i)\mapsto(e,i+1)$. In the lifted parity-check matrix the
shift $s(e)$ appears as the exponent of $x$, so that $\mathcal{G}^{(\ell)}$ is the
Tanner graph of a quasi-cyclic code with circulant blocks $P^{s(e)}$. We equip each lifted vertex
$(v,i)$ with the bijection $E((v,i))\to\{1,\dots,w\}$ inherited from $v$, so that
the Tanner code $\mathcal{T}(\mathcal{G}^{(\ell)},\mathcal{C}_0)$ is
$\mathbb{Z}_\ell$-invariant.

\begin{lemma}[{\cite{jeronimo2022explicit}}]
\label{lem:lift}
Let $\mathcal{G}$ be a $w$-regular graph on $2n_0$ vertices. There is a constant
$c_1$ and, for infinitely many $\ell$ with $\ell=2^{\Theta(n_0)}$, an explicitly
constructible (in polynomial time) $\ell$-cyclic lift $\mathcal{G}^{(\ell)}$ of
$\mathcal{G}$ with
\begin{equation}
  \lambda\bigl(\mathcal{G}^{(\ell)}\bigr)\ \le\ c_1\,\max\bigl(\lambda(\mathcal{G}),\sqrt{w}\bigr).
\end{equation}
\end{lemma}

For infinitely many degrees $w$ there exist $w$-regular Ramanujan graphs, with
$\lambda<2\sqrt{w-1}+1$ (see the references in \cite{jeronimo2022explicit}). Hence, as a corollary of
Lemma~\ref{lem:lift}, for such $w$ there exist $\mathbb{Z}_\ell$-lifts of $w$-regular base graphs
on $2n_0$ vertices with lift size $\ell=2^{\Theta(n_0)}$ and
$\lambda(\mathcal{G}^{(\ell)})\le c_2\sqrt{w}$ for an absolute constant $c_2$.
Choosing $w$ large enough that $c_2\sqrt{w}<\delta w$, Lemma~\ref{lem:expansion} applies to
$\mathcal{G}^{(\ell)}$ with $\alpha,\beta>0$ depending only on $w$ and $\delta$.

\begin{theorem}
\label{thm:h1}
Fix $\delta,\epsilon$ as in Lemma~\ref{lem:gv} and let $w$ be large enough for Lemma~\ref{lem:gv} and
for $c_2\sqrt{w}<\delta w$. Then there are constants $\alpha,\beta>0$ depending only
on $w,\delta$ such that, for infinitely many $\ell$, there exists a matrix
$H_1\in M_{m\times n}(R_\ell)$ with the following properties:
\begin{enumerate}
  \item $n=\Theta(\log\ell)$ and $\frac{m}{n}\in\bigl[\frac{\epsilon}{2},\frac{3\epsilon}{2}\bigr]$;
        in particular $m<n$ since $\epsilon<\frac23$;
  \item $H_1$ is $\max(w,2r)$-limited, where $r\le w$ is the number of rows of $H_0$;
  \item $\rho(H_1)$ and $\rho(H_1)^T=\rho(H_1^*)$ are both $(\alpha,\beta)$-expanding; equivalently (Remark~\ref{rem:transpose}), $\rho(H_1)$ and $\rho(H_1^T)$ are both $(\alpha,\beta)$-expanding.
\end{enumerate}
\end{theorem}

\begin{proof}
Let $\mathcal{C}_0$ be the code of Lemma~\ref{lem:gv}, with parity-check matrix $H_0$ whose
$r$ rows form a basis of $\mathcal{C}_0^\perp$, and let $\mathcal{G}$ be a
$w$-regular Ramanujan graph on $2n_0$ vertices. Let $\mathcal{G}^{(\ell)}$ be the
lift given by Lemma~\ref{lem:lift}, so that $\ell=2^{\Theta(n_0)}$ and
$\lambda(\mathcal{G}^{(\ell)})\le c_2\sqrt{w}<\delta w$. By Lemma~\ref{lem:expansion}, the canonical
parity-check matrix $\tilde H$ of $\mathcal{T}(\mathcal{G}^{(\ell)},\mathcal{C}_0)$,
of size $2n_0r\ell\times wn_0\ell$, and its transpose are both
$(\alpha,\beta)$-expanding, with $\alpha,\beta$ as in Lemma~\ref{lem:expansion} and depending only on
$w,\delta$.

Because the edge orderings at lifted vertices are inherited from the base graph,
the $\mathbb{Z}_\ell$-action on $\mathcal{G}^{(\ell)}$ permutes the rows and columns
of $\tilde H$ without changing its entries. Index the rows by $((v,j),i)$ and the
columns by $(e,i)$. Then $\tilde H$ is block-circulant: for a base edge $e=(u\to v)$
with shift $s(e)$ and a base vertex $v'$, the $\ell\times\ell$ block of $\tilde H$
with row index $(v',j)$ and column index $e$ equals
$(H_0)_{j,\,e}\,P^{0}$ if $v'=u$, $(H_0)_{j,\,e}\,P^{\,s(e)}$ if $v'=v$, and $0$
otherwise, where $(H_0)_{j,e}$ denotes the entry of $H_0$ in the column assigned to
$e$ at that vertex. Hence $\tilde H=\rho(H_1)$ for the matrix
$H_1\in M_{2n_0r\times wn_0}(R_\ell)$ with entries $(H_0)_{j,e}$ or
$(H_0)_{j,e}\,x^{s(e)}$, and $H_1$ is recovered from $\tilde H$ by reading the first
column of each block. Set $n=wn_0$ and $m=2n_0r$; then $n=\Theta(\log\ell)$ and
$m/n=2r/w\in[\epsilon/2,\,3\epsilon/2]$ by Lemma~\ref{lem:gv}, which gives property~1. Each row of $\tilde H$ has at most $w$ nonzero entries and each column at
most $2r$ (an edge meets two vertices, each carrying $r$ checks), which is
property~2. Finally $\rho(H_1)^T=\rho(H_1^*)$ by Eq.~\eqref{eq:rhostarT}, and both $\rho(H_1)$ and $\rho(H_1)^T=\tilde H^T$ are
$(\alpha,\beta)$-expanding, which is property~3.
\end{proof}

\begin{remark}
\label{rem:transpose}
Since $\rho(\bar H)=(I\otimes\Pi)\,\rho(H)\,(I\otimes\Pi)$ for the coordinate-reversal
permutation $\Pi$ on each $\ell$-block, expansion of $\rho(H_1)^T$ is equivalent to
expansion of $\rho(H_1^T)$, which is the form in which the hypothesis appears in
Lemmas~\ref{lem:case1}--\ref{lem:case2} and Theorem~\ref{thm:distance}.
\end{remark}

\subsection{Number of Logical Qudits of the Lifted Product Code}
\label{sec:lp-dimension}

\begin{theorem}[Lemma~1 in {\cite{panteleev2021quantum}}]
\label{thm:k}
Let $H_1\in M_{m\times n}(R_\ell)$ and let $b\in\mathbb{F}_q[x]$ be irreducible with
$b\mid x^\ell-1$, so that $R_{(b)}:=\mathbb{F}_q[x]/\langle b\rangle\cong\mathbb{F}_{q^{\deg b}}$.
Let $\beta\in\mathbb{F}_{q^{\deg b}}$ be a root of $b$ and write $H_1(\beta)$ for the
matrix over $\mathbb{F}_{q^{\deg b}}$ obtained by evaluating every entry of $H_1$ at
$x=\beta$. Then the fully stacked qudit lifted product code of
Definition~\ref{def:lp} with respect to $H_1$ and $H_2=b$ encodes
\begin{equation}
  k\ =\ \deg(b)\Bigl(\dim_{\mathbb{F}_{q^{\deg b}}}\ker H_1(\beta)
         +\dim_{\mathbb{F}_{q^{\deg b}}}\ker H_1(\beta)^{T}\Bigr),
\end{equation}
logical qudits. In particular $k\ge\deg(b)\,(n-m)$, and for the matrix $H_1$ of
Theorem~\ref{thm:h1} and $b=x-1$ one has $k=\Theta(n)=\Theta(\log\ell)=\Theta(\log N)$.
\end{theorem}

\begin{proof}
We include the proof for completeness. The fully stacked code is $\mathbb{F}_q$-linear (Corollary~\ref{cor:full-stacked-rate} and \cite{nadkarni2021mathbb}), so $k$ is the
$\mathbb{F}_q$-dimension of $\ker H_x/\operatorname{rowsp}(H_z)$, i.e.\
$k=N-\operatorname{rank}H_x-\operatorname{rank}H_z$ with $N=(n+m)\ell$. By Remark~\ref{rem:lp-signs}, $H_x=\rho([\,H_1\ \ bI_m\,])$ and
$H_z=\rho([\,\bar bI_n\ \ -H_1^*\,])$ up to the stacking factor, which does not change
ranks over $\mathbb{F}_q$. Using $\rho(M)^T=\rho(M^*)$,
\[
  \ker H_x^T=\{y\in R_\ell^{m}:\ H_1^*y=0,\ \bar b\,y=0\},\qquad
  \ker H_z^T=\{y\in R_\ell^{n}:\ b\,y=0,\ H_1y=0\}.
\]
Let $g=(x^\ell-1)/b$. Since $\operatorname{ann}_{R_\ell}(b)=\langle g\rangle$ and
$\operatorname{ann}_{R_\ell}(g)=\langle b\rangle$, the map $r\mapsto gr$ induces a
bijection $R_{(b)}\to\operatorname{ann}(b)$, and $H_1(g\tilde c)=g\,H_1(\beta)c$ for
any lift $\tilde c$ of $c\in R_{(b)}^{n}$. Hence $\ker H_z^T\cong\ker H_1(\beta)$ as
$\mathbb{F}_q$-spaces, and applying the antipode (a ring automorphism of $R_\ell$
exchanging $b$ with $\bar b$ and $H_1^*$ with $H_1^T$) gives
$\ker H_x^T\cong\ker H_1(\beta)^T$. Therefore
\[
  k=(n+m)\ell-\bigl(m\ell-\dim_{\mathbb{F}_q}\ker H_x^T\bigr)-\bigl(n\ell-\dim_{\mathbb{F}_q}\ker H_z^T\bigr)
   =\dim_{\mathbb{F}_q}\ker H_1(\beta)+\dim_{\mathbb{F}_q}\ker H_1(\beta)^T,
\]
and $\dim_{\mathbb{F}_q}=\deg(b)\cdot\dim_{\mathbb{F}_{q^{\deg b}}}$. The bound
$k\ge\deg(b)(n-m)$ follows from $\dim\ker H_1(\beta)\ge n-m$. For $b=x-1$,
$\deg b=1$ and $n-m=\Theta(n)$ by Theorem~\ref{thm:h1}, while $N=(n+m)\ell$ gives
$\log N=\Theta(n)$.
\end{proof}

\subsection{Minimum Distance of Lifted Product Code}
\label{sec:lp-distance}

Throughout this subsection, a vector $u\in\R^n$ is identified with its lift in $\F^{\ell n}$, and $\wt{u}$ denotes the Hamming weight of that lift; thus $\wt{fu}\le\wt{f}\,\wt{u}$ for $f\in\R$ and $\wt{x^tu}=\wt{u}$. For a vector $c$ over any ring, $\wtH{c}$ denotes the number of nonzero coordinates. A matrix $A\in M_{m\times n}(\R)$ is \emph{$w$-limited} if every row and every column of $\rho(A)$ has at most $w$ nonzero entries, so that $\wt{Ah}\le w\wt{h}$, and we write $\mathcal{C}(A):=\{x\in\R^n : Ax=0\}$. The matrix $A$ is said to be $(\alpha,\beta)$-expanding if $\rho(A)$ is, so that every nonzero $c\in\mathcal{C}(A)$ has $\wt{c}>\alpha\ell n$. We assume $0<\alpha<1$ and $0<\beta\le1$.

Fix $b\in\F[x]$ with $b(0)\ne0$, $b\notin\F^\times$, and let $e := \operatorname{ord}(b)$ be the least positive integer with $b\mid x^e-1$ (this exists since $b(0)\neq 0$, and does not require $\gcd(e,p)=1$). We take lift sizes $\ell$ with $e\mid\ell$, equivalently $b\mid x^\ell-1$, and set $L:=\ell/e$. Define in $\F[x]$
\begin{equation}\label{eq:qdef}
  q_1 := \frac{x^e-1}{b},\qquad N_e := \sum_{j=0}^{L-1}x^{je} = \frac{x^\ell-1}{x^e-1},\qquad
  g := \frac{x^\ell-1}{b} = q_1N_e,
\end{equation}
and for $t=1,\dots,L$
\begin{eqnarray}\label{eq:qt}
  q_t := \frac{x^{te}-1}{b} = q_1\sum_{j=0}^{t-1}x^{je}, &\qquad&
  b\,q_t = x^{te}-1,\nonumber\\
   q_{t+1}-q_t = x^{te}q_1,&\qquad& q_L = g .
\end{eqnarray}
Note $\wt{q_1}\le e$, $\wt{b}\le e+1$, and $\wt{x^{te}-1}=2$ in $\R$ for $0<te<\ell$.

\noindent {\bf Annihilator code and periodic vectors:}
$\R$ is a principal ideal ring, and for divisors $f\mid x^\ell-1$ one has $\ann_{\R}(f) = \bigl((x^\ell-1)/f\bigr)$. Hence
\[
  \Cb := \ann_{\R}(b) = (g), \qquad \ann_{\R}(g) = (b).
\]
Every $c\in\Cb$ is $c = gr = (q_1r)N_e$, the $L$-fold repetition of the length-$e$ word $q_1r\bmod(x^e-1)$. Let $\mathcal{P}:=(N_e)\subseteq\R$; its elements are described next.

Let $h\in\R$ be a single ring element, written through its coefficient vector
\[
  h=\sum_{j=0}^{\ell-1}c_j\,x^j,\qquad c_j\in\F .
\]
Since $\ell=eL$, every exponent $j\in\{0,\dots,\ell-1\}$ has a unique decomposition $j=\rho+ke$ with $\rho\in\{0,\dots,e-1\}$ and $k\in\{0,\dots,L-1\}$. For each residue $\rho$ define
\begin{equation}\label{eq:hrho}
  h_\rho:=\bigl(c_\rho,\ c_{\rho+e},\ c_{\rho+2e},\ \dots,\ c_{\rho+(L-1)e}\bigr)\in\F^{L},
\end{equation}
the vector of coefficients of $h$ at exponents congruent to $\rho$ modulo $e$, listed in increasing order. Equivalently, as a polynomial in a new variable $y$,

\begin{eqnarray*}
  h_\rho(y) &=& \sum_{k=0}^{L-1}c_{\rho+ke}\,y^k\in\F[y]/(y^L-1),\\
  h& =&\sum_{\rho=0}^{e-1}x^\rho\,h_\rho(x^e).
\end{eqnarray*}
The map $h\mapsto(h_0,\dots,h_{e-1})$ is an $\F$-linear, weight-preserving bijection $\R\to(\F^L)^e$. Two facts follow from \eqref{eq:hrho}.
\begin{enumerate}
  \item Multiplication by $x^e$ moves $c_j$ to exponent $j+e$ (mod $\ell$). On the class $\rho$ this is $k\mapsto k+1$ (mod $L$). Hence $(x^{te}h)_\rho=y^t\,h_\rho$, the cyclic shift by $t$ positions inside $\F^L$, for every $t$ and $\rho$.
  \item $x^\rho N_e=\sum_{k=0}^{L-1}x^{\rho+ke}$ is the element whose class $\rho$ is the all-ones vector and whose other classes are zero. Since $x^jN_e=x^{j\bmod e}N_e$, the ideal $\mathcal{P}=(N_e)$ is the $\F$-span of $x^0N_e,\dots,x^{e-1}N_e$. Therefore, 
    $h\in \mathcal{P}\iff h_\rho\ \text{is a constant vector } (c,c,\dots,c)\ \text{for every } \rho\in\{0,\dots,e-1\}$.
\end{enumerate}
We call $\mathcal{P}$ the ideal of $e$-periodic elements. Then $\Cb\subseteq \mathcal{P}$, $x^ec=c$ for $c\in \mathcal{P}$, and every nonzero $c\in \mathcal{P}$ has $\wt{c}\ge L$ (a nonzero constant class contributes $L$).

For $y\in\F^L$ let $M(y):=\max_{c\in\F}\#\{k : y_k = c\}$ be the largest multiplicity of a value among its coordinates. The distance of a ring element $h$ to the ideal of $e$-periodic elements is then
\begin{equation}\label{eq:dist}
  \dist(h,\mathcal{P}) := \min_{\pi\in \mathcal{P}}\wt{h-\pi} = \sum_{\rho=0}^{e-1}\bigl(L - M(h_\rho)\bigr),
\end{equation}
since the nearest constant vector to $h_\rho$ is at Hamming distance $L-M(h_\rho)$. For a vector $h=(h_1,\dots,h_n)\in\R^n$, with $h_{i,\rho}$ denoting class $\rho$ of the coordinate $h_i$, the distance of $h$ to $\mathcal{P}^n$ is
\begin{equation} \label{eq:delta}
\delta(h):=\sum_{i=1}^{n}\dist(h_i,\mathcal{P}) = \sum_{i=1}^{n}\sum_{\rho=0}^{e-1}\bigl(L-M(h_{i,\rho})\bigr) = \min_{\pi\in\mathcal{P}^n}\wt{h-\pi}.
\end{equation}
Let $\Rb:=\F[x]/(b)$ and $\varphi:\R\to\Rb$ the reduction map, applied entrywise, and put $B:=\varphi(A)\in M_{m\times n}(\Rb)$. Since $gb=0$, $g\,r$ depends only on $\varphi(r)$, and the induced map
\begin{equation}\label{eq:gmap}
  \Rb\longrightarrow\Cb,\qquad \varphi(r)\longmapsto gr,
\end{equation}
is injective (as $\ann(g)=(b)$), hence bijective. For $c\in\Rb^n$ with any lift $\tilde c\in\R^n$ one has $A(g\tilde c)=g\,Bc$.

We now consider the lifted product code of Definition~\ref{def:lp} with $H_1=A$ and $H_2=b$, in the sign convention of Remark~\ref{rem:lp-signs}: $H_x=[A,\,-bI_m]$, $H_z=[\bar bI_n,\,A^*]$ (as matrices over $R_\ell$). We write $\CZ:=\{z\in\R^{n+m}: H_xz^T=0\}$ for the $Z$-type normalizer (this is the code $\mathcal{A}_x$ of Sec.~\ref{sec:additive-codes}, since the code is fully stacked) and $\CXp$ for the $\F$-row space of $H_z$, i.e.\ the $Z$-type stabilizers. For $z=[u,v]$ with $u\in\R^n$, $v\in\R^m$,
\begin{eqnarray}\label{eq:CZ}
  &&z\in\CZ\iff Au=bv,\\ &&z\in\CXp\iff \exists h\in\R^n:\ u=bh,\ v=Ah .
\end{eqnarray}

\begin{lemma}[Case 1]
\label{lem:case1}
Let $A\in M_{m\times n}(\R)$ be $w$-limited with $A$ $(\alpha,\beta)$-expanding, and let $z=[u,v]\in\CZ$ with $\varphi(u)\ne0$. Then
\[
  \wt{u}+\wt{v}\ \ge\ \gamma_1\,\ell n,\qquad \gamma_1:=\min\Bigl(\frac{\alpha}{2e},\ \frac{\alpha\beta}{4}\Bigr).
\]
\end{lemma}

\begin{proof}
The idea is to apply the expansion property of $A$ not to $u$ itself but to the multiples $q_tu$, one of which has weight in the range where expansion applies.

From \eqref{eq:CZ}, $Au=bv$, so $A(gu)=g\,Au=g\,b\,v=0$ because $gb=x^\ell-1=0$ in $\R$. Thus $gu\in\mathcal{C}(A)$, and $gu\ne0$ by \eqref{eq:gmap} since $\varphi(u)\ne0$. Expansion gives directly
\begin{equation}\label{eq:theta}
  \wt{gu}>\theta:=\alpha\ell n .
\end{equation}
If $\wt{u}>\theta/(2e)$, then $\wt{z}\ge\wt{u}>\alpha\ell n/(2e)\ge\gamma_1\ell n$ and we are done. Thus henceforth, we assume that $\wt{u}\le\theta/(2e)$. For $t=1,\dots,L$ set $u^{(t)}:=q_tu\in\R^n$. By \eqref{eq:qt},

\begin{eqnarray*} && \wt{u^{(1)}} \le \wt{q_1}\wt{u}\le e\wt{u}\le\frac{\theta}{2},\\
  && \bigl|\wt{u^{(t+1)}}-\wt{u^{(t)}}\bigr|\le\wt{x^{te}q_1u}  \le e\wt{u}\le\frac{\theta}{2}, \\
 && \wt{u^{(L)}}=\wt{gu}>\theta.
\end{eqnarray*}

A sequence that starts at or below $\theta/2$, ends above $\theta$, and moves by at most $\theta/2$ per step must take a value in $[\theta/2,\theta]$. Fix $t_0$ with
\begin{equation}\label{eq:window}
  \frac{\theta}{2}\le\wt{u^{(t_0)}}\le\theta=\alpha\ell n .
\end{equation}
Using $Au=bv$ and \eqref{eq:qt},
$Au^{(t_0)}=q_{t_0}Au=q_{t_0}b\,v=(x^{t_0e}-1)v$. As a result, we have $ \wt{Au^{(t_0)}}\le2\wt{v}$.
By \eqref{eq:window}, $u^{(t_0)}$ lies in the domain where expansion applies, so
\[
  2\wt{v}\ \ge\ \wt{Au^{(t_0)}}\ \ge\ \beta\wt{u^{(t_0)}}\ \ge\ \frac{\beta\theta}{2}\ =\ \frac{\alpha\beta\,\ell n}{2}.
\]
Therefore $\wt{z}\ge\wt{v}\ge\frac{\alpha\beta}{4}\ell n\ge\gamma_1\ell n$.
\end{proof}

\begin{lemma}[Structure of Case 2]\label{lem:structure}
Let $z=[u,v]\in\CZ\setminus\CXp$ with $\varphi(u)=0$. Then there are $h\in\R^n$ and $v'\in\Rb^m$ with
\begin{equation}\label{eq:case2form}
  u=bh,\qquad v=Ah+gv',\qquad v'\notin\im B .
\end{equation}
In particular $v'\ne0$. The pair $(h,v')$ is unique up to $(h,v')\mapsto(h+g\tilde s,\ v'-Bs)$, $s\in\Rb^n$.
\end{lemma}
\begin{proof}
$\varphi(u)=0$ means $u\in(b)^n$, so $u=bh$ for some $h$. Then $b(Ah-v)=Au-bv=0$ by \eqref{eq:CZ}, so $Ah-v\in\Cb^m=(g)^m$, i.e.\ $v=Ah+gv'$ for a unique $v'\in\Rb^m$ by \eqref{eq:gmap}. If $v'=Bs$ for some $s\in\Rb^n$ with lift $\tilde s$, then $gv'=gBs=A(g\tilde s)$ and $b\,g\tilde s=0$, so $z=[b(h+g\tilde s),\,A(h+g\tilde s)]\in\CXp$ by \eqref{eq:CZ}, contradicting nondegeneracy. Hence $v'\notin\im B$, and $v'\ne0$ since $0=B\cdot0\in\im B$. The freedom in $h$ is $\ker(b)^n=\Cb^n=(g\tilde s)$, which shifts $v'$ by $-Bs$.
\end{proof}

\begin{lemma}[Translate lemma]\label{lem:translate}
For every $h\in\R^n$ there is $t\in\{1,\dots,L-1\}$ with $\wt{(x^{te}-1)h}\ge\delta(h)$.
\end{lemma}
\begin{proof}
Since $x^{te}$ shifts each class by $t$ and the class decomposition preserves weight,
$\wt{(x^{te}-1)h}=\sum_{i,\rho}\wt{(y^t-1)h_{i,\rho}}$, where $y$ denotes the cyclic shift on $\F^L$. For a single $y\in\F^L$ with coordinates $y_0,\dots,y_{L-1}$, the $k$-th coordinate of $(y^t-1)y$ is $y_{k-t}-y_k$ (indices mod $L$), nonzero iff $y_k\ne y_{k-t}$. With $m_c:=\#\{k:y_k=c\}$,
\begin{eqnarray*}
  \sum_{t=0}^{L-1}\wt{(y^t-1)y} & = & \#\{(k,t):y_k\ne y_{k-t}\} \\
  & = &L^2-\sum_{c\in\F}m_c^2\ \\ &\ge & L^2-M(y)\sum_cm_c=L\,(L-M(y)),
\end{eqnarray*}
using $m_c\le M(y)$ and $\sum_cm_c=L$. The $t=0$ term vanishes. Summing over $i$ and $\rho$ and using \eqref{eq:delta}, $\sum_{t=1}^{L-1}\wt{(x^{te}-1)h}\ge L\,\delta(h)$, so some $t\ge1$ attains at least $L\delta(h)/(L-1)\ge\delta(h)$.
\end{proof}

\begin{lemma}[Distance of $h$ from $\mathcal{P}^n$]\label{lem:far}
Let $z=[u,v]\in\CZ\setminus\CXp$ with $\varphi(u)=0$, written as in \eqref{eq:case2form}. Then
\[
  w'\,\delta(h)\ \ge\ L-\wt{z},\qquad w'=\max(w,\wt{b}).
\]
\end{lemma}
\begin{proof}
Write $h=\pi+\eta$ with $\pi\in \mathcal{P}^n$ and $\wt{\eta}=\delta(h)$, by \eqref{eq:delta}. Since $\mathcal{P}$ is an ideal, $A\pi\in \mathcal{P}^m$ and $b\pi\in \mathcal{P}^n$; also $gv'\in\Cb^m\subseteq \mathcal{P}^m$. From \eqref{eq:case2form},
\[
  A\pi+gv'=v-A\eta\in \mathcal{P}^m,\qquad b\pi=u-b\eta\in \mathcal{P}^n .
\]
If $A\pi+gv'\ne0$, it is a nonzero periodic vector, so $L\le\wt{v-A\eta}\le\wt{v}+w\,\delta(h)$. If $b\pi\ne0$, likewise $L\le\wt{u-b\eta}\le\wt{u}+\wt{b}\,\delta(h)$. In either case $L\le\wt{z}+w'\delta(h)$.

It remains to exclude $A\pi=-gv'$ and $b\pi=0$ together. Then $\pi\in\ann(b)^n=\Cb^n$, so $\pi=g\tilde s$ for some $s\in\Rb^n$, and $gv'=A\pi=A(g\tilde s)=g\,Bs$; injectivity of \eqref{eq:gmap} gives $v'=Bs\in\im B$, contradicting Lemma~\ref{lem:structure}.
\end{proof}

\begin{lemma}[Case 2]\label{lem:case2}
Let $A\in M_{m\times n}(\R)$ be $w$-limited with $A$ $(\alpha,\beta)$-expanding. Let $z=[u,v]\in\CZ\setminus\CXp$ with $\varphi(u)=0$. Then
\[
  \wt{u}+\wt{v}\ \ge\ \gamma_2\,\ell,\qquad \gamma_2:=\frac{\alpha\beta}{4ew'} .
\]
\end{lemma}

\begin{proof}
Write $z$ as in \eqref{eq:case2form} and suppose, for a contradiction, that $\wt{z}<\gamma_2\ell$; then also $\wt{u}<\gamma_2\ell$ and $\wt{v}<\gamma_2\ell$.

\noindent {\bf Step 1: $h$ is far from periodic.}
\noindent Since $\alpha,\beta\le1$ and $w'\ge1$, $\gamma_2\le\frac{1}{4e}$, so $\wt{z}<L/4$. Lemma~\ref{lem:far} gives
\[
  \delta(h)\ \ge\ \frac{L-\wt{z}}{w'}\ \ge\ \frac{3L}{4w'}\ =\ \frac{3\ell}{4ew'} .
\]
Set $\tau:=\dfrac{\alpha\ell}{2ew'}$. Because $\alpha<1$, $\tau<\frac{\ell}{2ew'}\le\frac{3\ell}{4ew'}\le\delta(h)$.

\noindent {\bf Step 2: Translates of $h$.}
For $1\le t\le L-1$ put $h_t:=(x^{te}-1)h$ and $W_t:=\wt{h_t}$, with $W_0:=0$. By \eqref{eq:qt} and $u=bh$,
\[
  h_1=(x^e-1)h=q_1bh=q_1u,\qquad
  h_t=q_1u+x^e\,h_{t-1}\quad(t\ge2),
\]
the second from $x^{te}-1=(x^e-1)+x^e(x^{(t-1)e}-1)$. Hence $W_t\le W_{t-1}+e\wt{u}<W_{t-1}+e\gamma_2\ell$ for all $t\ge1$. Lemma~\ref{lem:translate} and Step 1 give $t'$ with $W_{t'}\ge\delta(h)\ge\tau$. Let $t_0\ge1$ be the least index with $W_{t_0}\ge\tau$. Then $W_{t_0-1}<\tau$, so
\begin{eqnarray}\label{eq:window2}
  \tau\ & \le&\ W_{t_0}\ <\ \tau+e\gamma_2\ell\ =\ \frac{\alpha\ell}{2ew'}+\frac{\alpha\beta\ell}{4w'} \\
  &\le &\alpha\ell\Bigl(\frac12+\frac14\Bigr)\ <\ \alpha\ell\ \le\ \alpha\ell n .
\end{eqnarray}
This places $h_{t_0}$ in the expansion domain.

\noindent {\bf Step 3: Syndrome of $h_{t_0}$ is controlled by $v$.}
\noindent Since $g\in \mathcal{P}$, $x^{t_0e}g=g$, so $(x^{t_0e}-1)gv'=0$ and by \eqref{eq:case2form}
\begin{eqnarray*}
  && Ah_{t_0}=(x^{t_0e}-1)Ah=(x^{t_0e}-1)(v-gv')=(x^{t_0e}-1)v, \\
  && \implies \wt{Ah_{t_0}}\le2\wt{v}<2\gamma_2\ell.
\end{eqnarray*}

\noindent {\bf Step 4: Application of expansion property.}
\noindent By \eqref{eq:window2},
\[
  \wt{Ah_{t_0}}\ \ge\ \beta W_{t_0}\ \ge\ \beta\tau\ =\ \frac{\alpha\beta\ell}{2ew'}\ =\ 2\gamma_2\ell ,
\]
contradicting Step 3. Hence $\wt{z}\ge\gamma_2\ell$.
\end{proof}

\begin{theorem}[Minimum distance of $\LP(A,b)$ over $\F$]\label{thm:distance}
Let $q$ be a prime power, $\ell\ge2$, and let $b\in\F[x]$ with $b(0)\ne0$, $b\notin\F^\times$,
of order $e=\operatorname{ord}(b)$ with $e\mid\ell$. Let $A\in M_{m\times n}(\R)$ be $w$-limited
with both $A$ and $A^{T}$ $(\alpha,\beta)$-expanding, $0<\alpha<1$, $0<\beta\le1$, and put
$w':=\max(w,\wt{b})$. Then the quantum code $\mathcal{Q}=\LP(A,b)$ of length $N=(n+m)\ell$
over $\F$ satisfies
\[
  d(\mathcal{Q})\ \ge\ \gamma\,\ell,\qquad
  \gamma:=\min\Bigl(\frac{\alpha}{2e},\ \frac{\alpha\beta}{4ew'}\Bigr),
\]
with no further condition on $n$, $m$ or $\ell$. More precisely:
\begin{enumerate}
  \item $d_Z(\mathcal{Q})\ \ge\ \min\bigl(\gamma_1\ell n,\ \gamma_2\ell\bigr)$ and
        $d_X(\mathcal{Q})\ \ge\ \min\bigl(\gamma_1\ell m,\ \gamma_2\ell\bigr)$, where
        $\gamma_1=\min\bigl(\frac{\alpha}{2e},\frac{\alpha\beta}{4}\bigr)$ and
        $\gamma_2=\frac{\alpha\beta}{4ew'}$.
  \item If $B=\varphi(A)$ is surjective onto $\Rb^{m}$ then $d_Z(\mathcal{Q})\ge\gamma_1\ell n$;
        if $B^{T}$ is surjective onto $\Rb^{n}$ then $d_X(\mathcal{Q})\ge\gamma_1\ell m$.
  \item If $m<n$ then $d_X(\mathcal{Q})\le\wt{g}\le\ell$; if $n<m$ then $d_Z(\mathcal{Q})\le\wt{g}\le\ell$.
        In particular $d(\mathcal{Q})=\Theta(\ell)$ whenever $m\ne n$.
\end{enumerate}
For $b=x-1$ the constants are $e=1$, $w'=\max(w,2)$, $\gamma_1=\alpha\beta/4$,
$\gamma_2=\alpha\beta/(4w')$, hence $d(\mathcal{Q})\ge\frac{\alpha\beta}{4w'}\,\ell$.
\end{theorem}

\begin{proof}
\emph{$Z$-distance.} Let $z=[u,v]\in\CZ\setminus\CXp$. Either $\varphi(u)\ne0$, and
Lemma~\ref{lem:case1} gives $\wt{z}\ge\gamma_1\ell n$; or $\varphi(u)=0$, and
Lemma~\ref{lem:case2} gives $\wt{z}\ge\gamma_2\ell$. Hence
$d_Z\ge\min(\gamma_1\ell n,\gamma_2\ell)$. If $B$ is surjective, Lemma~\ref{lem:structure}
shows that no nondegenerate $z$ with $\varphi(u)=0$ exists (it would need
$v'\notin\im B=\Rb^{m}$), so only Case~1 occurs and $d_Z\ge\gamma_1\ell n$.

\emph{$X$-distance.} $X$-vectors $[u,v]$ satisfy $H_z[u,v]^{*}=0$, i.e.\ $\bar b u+A^{*}v=0$,
and the degenerate ones are the $\F$-combinations of the columns of
$H_x^{*}=\begin{bmatrix}A^{*}\\ -\bar b I_m\end{bmatrix}$, i.e.\ $[A^{*}h,\,-\bar b h]$.
The weight-preserving substitution $(u,v)\mapsto(v,-u)$ turns these into
$A^{*}v=\bar b u$ and $[\bar b h,\,A^{*}h]$, which is the $Z$-code of $\LP(A^{*},\bar b)$.
The antipode $x\mapsto x^{-1}$ permutes the coordinates of each $\ell$-block, so it
preserves weights and maps $\LP(A^{*},\bar b)$ onto $\LP(A^{T},b)$. Hence
$d_X(\LP(A,b))=d_Z(\LP(A^{T},b))$. Since $A^{T}$ is $w$-limited and $(\alpha,\beta)$-expanding,
and $\operatorname{ord}(b)$, $\wt{b}$ are unchanged, the $Z$-argument applied to $A^{T}$
(with $n\leftrightarrow m$ and $B\leftrightarrow B^{T}$) gives
$d_X\ge\min(\gamma_1\ell m,\gamma_2\ell)$, and $d_X\ge\gamma_1\ell m$ when $B^{T}$ is surjective.

\emph{Overall bound.} $\gamma_2\le\alpha\beta/4$ and $\ell n,\ell m\ge\ell$, so
$\min(\gamma_1\ell n,\gamma_2\ell)\ge\min(\gamma_1,\gamma_2)\ell=\gamma\ell$, and likewise for $d_X$.

\emph{Upper bound.} If $m<n$ then $|\Rb^{m}|<|\Rb^{n}|$, so $B^{T}:\Rb^{m}\to\Rb^{n}$ is not
surjective and some standard basis vector $e_i\notin\im B^{T}$. By Lemma~\ref{lem:structure}
applied to $\LP(A^{T},b)$, $[0,\,g e_i]$ is a nondegenerate $Z$-codeword of $\LP(A^{T},b)$,
of weight $\wt{g}\le\ell$. Hence $d_X(\LP(A,b))=d_Z(\LP(A^{T},b))\le\wt{g}$.
The case $n<m$ is symmetric.
\end{proof}

Combining Theorems~\ref{thm:h1}, \ref{thm:k} and \ref{thm:distance} gives the parameters of the explicit family.

\begin{corollary}
\label{cor:main-params}
Fix $\delta,\epsilon$ and $w$ as in Theorem~\ref{thm:h1}, and let $H_1$ be the matrix of Theorem~\ref{thm:h1}, with $w_1:=\max(w,2r)$. Then the fully stacked qudit lifted product code with respect to $H_1$ and $H_2=x-1$ is a QLDPC code over $\F$ of length $N=(n+m)\ell$ with
\[
  k\ \ge\ n-m\ =\ \Theta(\log N),\qquad
  d\ \ge\ \frac{\alpha\beta}{4\max(w_1,2)}\,\ell\ =\ \Theta\Bigl(\frac{N}{\log N}\Bigr),
\]
where $\alpha,\beta$ depend only on $w$ and $\delta$.
\end{corollary}
\begin{proof}
By Theorem~\ref{thm:h1}, $H_1$ is $w_1$-limited, $\rho(H_1)$ and $\rho(H_1^T)$ are $(\alpha,\beta)$-expanding (Remark~\ref{rem:transpose}), and $m<n$ with $n=\Theta(\log\ell)$. Theorem~\ref{thm:k} with $b=x-1$ gives $k\ge n-m=\Theta(n)$, and Theorem~\ref{thm:distance} with $e=1$, $\wt{b}=2$ gives $d\ge\frac{\alpha\beta}{4\max(w_1,2)}\ell$. Since $N=(n+m)\ell$ with $n+m=\Theta(\log\ell)$, we have $\log N=\Theta(\log \ell)=\Theta(n)$ and $\ell=\Theta(N/\log N)$. Each row and column of $H_x$ and $H_z$ has at most $w_1+2$ nonzero entries, so the family is LDPC.
\end{proof}

\section{Decoding Strategies}
\label{sec:decoder}
We now proceed to the task of outlining decoding strategies for qudit QLDPC codes. We shall first begin with a brief review of the framework of iterative message passing decoders which are commonly used for the decoding of classical LDPC codes as well as quantum LDPC codes in the context of qubit systems. We shall then discuss two different strategies for the decoding of qudit QLDPC codes and comment on the advantages and disadvantages of either of these approaches.

\subsection{Iterative Decoding Framework}
\label{sec:decoder:msg_passing}

Iterative message-passing decoders operate by exchanging probabilistic information between the variable and check nodes of the Tanner graph associated with a code. A typical decoder proceeds through four stages:
\begin{itemize}
    \item Initialization.
    \item Variable node update.
    \item Check node update.
    \item Hard decision.
\end{itemize}
During initialization, each variable node is assigned probabilistic information obtained from the underlying noise model. The variable- and check-node update steps then iteratively refine these beliefs by exchanging messages along the edges of the Tanner graph. After each iteration, a hard-decision estimate of the error is computed. Decoding terminates successfully if this estimate satisfies the measured syndrome; otherwise, the iterations continue until either convergence or a maximum iteration count is reached. In this work, we assume a qudit depolarizing channel over $GF(q)$ as our noise model for decoder benchmarking.

\begin{definition}
    \label{def:depol-model}
    A qudit depolarizing channel over $GF(q)$ with error rate $\epsilon$ is defined as a $q$-ary symmetric channel in which the identity error occurs with probability $1-\epsilon$, while each non-trivial Pauli error $E \in \mathcal{G}_q \backslash \{I\}$ occurs with an identical probability of $\epsilon/(q^2-1)$.
\end{definition}

Following the code constructions introduced earlier, we further assume that the decoded error correction code is a CSS code, so that $X$ and $Z$ type errors may be corrected independently using the parity-check matrices $H_z$ and $H_x$, respectively. In this setting, the channel model may be simplified by considering only one error type at a time. Specifically, when decoding $Z$ type errors using $H_x$, each non-trivial $Z$ error occurs with probability $p/(q-1)$ and the identity error occurs with probability $1-p$, where the error rate $p$ is related to the qudit depolarizing rate $\epsilon$ as follows.
\begin{equation}
    \label{eq:depol-prob}
    p = \frac{q\epsilon}{q+1}.
\end{equation}

The decoding of $X$ type errors using $H_z$ proceeds analogously. The initialization step therefore consists of assigning to each variable node a probability distribution over possible errors. In binary decoders, this information is typically represented using a scalar log-likelihood ratio (LLR). In the non-binary setting, this generalizes naturally to an LLR vector.

\begin{definition}
    Let $\zeta$ be a primitive element of $GF(q)$ such that
    \begin{equation}
        GF(q)=\{0,1,\zeta,\zeta^2,\dots,\zeta^{q-2}\}.
    \end{equation}
    The LLR vector corresponding to a probability distribution $P(e)$ over $GF(q)$ is the vector $\Lambda$ defined by
    \begin{equation}
        \Lambda^{(i)} = \ln\frac{P(0)}{P(\zeta^i)}.
    \end{equation}
\end{definition}

We denote the variable-to-check message sent from variable node $v_n$ to check node $c_m$ by $\Gamma_{n \to m}$, and the corresponding check-to-variable message by $\Delta_{m \to n}$. During decoding, variable nodes update outgoing messages using the channel prior together with incoming messages from neighboring check nodes, while check nodes update outgoing messages using the measured syndrome and incoming messages from neighboring variable nodes. In both cases, the outgoing message along a given edge is computed independently of the previous message received along that same edge.

The precise forms of these update rules depend on the decoder under consideration. In the following sections, we discuss two message-passing decoders for qudit QLDPC codes. The first is a direct generalization of the decoder proposed in~\cite{lai2021log} and applies to arbitrary qudit QLDPC codes. The second decoder is only applicable to fully stacked QLDPC codes but achieves a significantly improved decoding performance in those cases.

\subsection{Decoding Trace Additive Codes}
\label{sec:decoder:additive}
We now describe the process of decoding trace additive qudit codes over $GF(q)$ via belief propagation. An important caveat of decoding trace additive codes is that while the error vector $e$ for a code over $GF(q)$ has its elements lying in $GF(q)$, the corresponding syndrome $s$ lies in the base field $GF(p)$, where $q = p^m$. This is because the syndrome is evaluated in accordance with the equation,
\begin{equation}
    s = tr(e H^T),
\end{equation}
where $tr:GF(q) \to GF(p)$ represents the field trace defined in Def.~\ref{def:trace}. This creates a mismatch between the operating fields of the variable and check node update operations, since the check nodes must now operate over $GF(p)$ whereas the variable nodes must operate over $GF(q)$. To handle this mismatch, we follow the same approach as introduced in~\cite{lai2021log} and introduce a belief quantization function that translates LLR vectors over $GF(q)$ to LLR vectors over $GF(p)$.
\begin{definition}
    \label{def:belief_quant}
    Let $\Lambda_n$ be a $q$-ary LLR vector obtained as a variable-to-check message $\Gamma_{n \to m}$ on an edge $v_n \to c_m$. Let $\eta \in GF(q)$ be the value of the corresponding edge in the associated parity check matrix. We define the belief quantization function $\lambda_\eta : \mathbb{R}^{q-1} \to \mathbb{R}^{p-1}$ as follows.
    \begin{equation}
    \begin{split}
        \lambda_\eta(\Lambda_n)^{(j)} &= \ln\frac{\sum_{\xi \in GF(q) : \expval{\xi, \eta} = 0} \text{\emph{Pr}}(e_n = \xi)}{\sum_{\xi \in GF(q) : \expval{\xi, \eta} = j} \text{\emph{Pr}}(e_n = \xi)} \\
        &= \ln\frac{1 + \sum_{i : \expval{\zeta^i, \eta} = 0} e^{-\Lambda_n^{(i)}}}{\sum_{i : \expval{\zeta^i, \eta}=j} e^{-\Lambda_n^{(i)}}},
    \end{split}
    \end{equation}
    where $j \in GF(p)\backslash\{0\}$, $e_n \in GF(q)$ is the error symbol on the node $v_n$ and $\expval{a, b} = tr(ab)$ represents the trace inner product. 
\end{definition}

From the above definition, we can see that physically, the $j^{th}$ term of the belief quantization function output essentially evaluates the following LLR.
\begin{equation}
    \lambda_\eta(\Lambda_n)^{(j)} = \ln\frac{\text{Pr}(\expval{e_n, \eta} = 0)}{\text{Pr}(\expval{e_n, \eta} = j)}.
\end{equation}
Hence, belief quantization achieves the task of translating a $q$-ary LLR into a $p$-ary LLR. For each edge $v_n \to c_m$, we may use belief quantization on the variable-to-check messages to translate $q$-ary LLRs into $p$-ary LLRs. The check node operations can then expect $p$-ary LLRs as input and operate over $GF(p)$. The check node operation itself is typically an FFT operation over the input $p$-ary LLRs that produces $p$-ary LLRs as output check-to-variable messages~\cite{zhang2016vlsi}. These $p$-ary LLRs must now be accumulated via the variable node operation to produce $q$-ary LLRs over the possible errors in $GF(q)$ that can then be used for hard-decision and subsequent variable-to-check messages. Algorithm~\ref{algo:refined_bp} shows the full set of operations to be performed.
\begin{algorithm}[t]
\caption{BP Decoding of Qudit QLDPC Codes Over $\mathrm{GF}(q)$}
\label{algo:refined_bp}
\begin{algorithmic}[1]
\State \textbf{Input:} Parity check matrix $H \in \mathrm{GF}(q)^{M\times N}$, target syndrome $s \in \mathrm{GF}(p)^M$, max iterations $T_{\max}\in\mathbb{Z}^+$, and channel
LLR vectors $\{\Lambda_n \in \mathbb{R}^{q-1}\}_{n=1}^N$.
\State \textbf{Initialization:} Set $\Gamma_{n\to m} = \Lambda_n$ for all edges $(m, n)$.
\Repeat
\State \textbf{Horizontal step:} For $m\in [M]$ and $n\in\mathcal N(m)$:
\For{$n' \in \mathcal N(m)\setminus n$}
\State Set $\mu_{n'}(y) = \exp\big(-\lambda_{H_{mn'}}(\Gamma_{n'\to m})(y)\big)$, \item[] \qquad\quad $y\in\mathrm{GF}(p)$, with $\lambda_{H_{mn'}}(\Gamma_{n'\to m})(0) = 0$.
\EndFor
\State Compute
\[
\tau_{m\to n}(y) = \mathcal{F}^{-1}\left[\chi_{s_m} \cdot \prod_{n'\in\mathcal N(m)\setminus n} \mathcal{F}^*[\mu_{n'}] \right](y),
\]
\quad\, where $\mathcal{F}$ is the discrete Fourier transform over $\mathrm{GF}(p)$, \item[] \quad\, and $\chi_{s_m}(\xi) = \exp(-2\pi i \; \xi s_m / p)$.
\State Set, for $j \in \mathrm{GF}(p)\setminus\{0\}$,
\[
\Delta_{m\to n}^{(j)} = \ln \frac{\tau_{m\to n}(0)}{\tau_{m\to n}(j)}, \qquad \Delta_{m\to n}^{(0)} = 0.
\]
\State \textbf{Vertical step:} For $n \in [N]$, $m\in\mathcal N(n)$ and $i = 0,\dots,q-2$, compute
\[
\Gamma_{n\to m}^{(i)} = \Lambda_n^{(i)} + \sum_{\substack{m'\in\mathcal N(n)\setminus m \\ j = tr(H_{m'n}\zeta^i)}} \Delta_{m'\to n}^{(j)}.
\]
\State \textbf{Hard-decision step:} For $n \in [N]$, compute
\[
\hat\Lambda_n^{(i)} = \Lambda_n^{(i)} + \sum_{\substack{m\in\mathcal N(n) \\ j = tr(H_{mn}\zeta^i)}} \Delta_{m\to n}^{(j)},
\]
\quad\, and set
\[
\hat e_n = \begin{cases} 0, & \hat\Lambda_n^{(i)} > 0\ \ \forall i, \\ \zeta^{\arg\min_i \hat\Lambda_n^{(i)}}, & \text{otherwise.} \end{cases}
\]
\Until{$tr(\hat e \cdot H^T) = s$ \textbf{or} $T_{\max}$ iterations are reached.}
\end{algorithmic}
\end{algorithm}

\begin{algorithm}[t]
\caption{BP Decoding of Fully Stacked Qudit QLDPC Codes Over $\mathrm{GF}(q)$ With Merged Syndrome}
\label{algo:linear_bp}
\begin{algorithmic}[1]
\State \textbf{Input:} Unstacked parity check matrix $H \in \mathrm{GF}(q)^{M\times N}$ ($H_{x0}$ or $H_{z0}$), merged syndrome $\tilde s \in \mathrm{GF}(q)^M$ (Eq.~\ref{eq:merge-syn}), max iterations $T_{\max}\in\mathbb{Z}^+$, and channel
LLR vectors $\{\Lambda_n \in \mathbb{R}^{q-1}\}_{n=1}^N$.
\State \textbf{Initialization:} Set $\Gamma_{n\to m} = \Lambda_n$ for all edges $(m, n)$.
\Repeat
\State \textbf{Horizontal step:} For $m\in [M]$ and $n\in\mathcal N(m)$:
\For{$n' \in \mathcal N(m)\setminus n$}
\State Set $\kappa_{n'}(y) = \exp\big(-\Gamma_{n'\to m}(H_{mn'}^{-1}y)\big)$, \item[] \qquad\quad $y\in\mathrm{GF}(q)$, with $\Gamma_{n'\to m}(0) = 0$.
\EndFor
\State Compute
\[
\tau_{m\to n}(y) = \mathcal{F}^{-1}\left[\chi_{\tilde s_m} \cdot \prod_{n'\in\mathcal N(m)\setminus n} \mathcal{F}^*[\kappa_{n'}] \right](y),
\]
\quad\, where $\mathcal{F}$ is the discrete Fourier transform over $\mathrm{GF}(q)$, \item[] \quad\, and $\chi_{\tilde s_m}(\xi) = \exp(-2\pi i \;tr(\xi \tilde{s}_m) / p)$.
\State Set, for $i = 0,\dots,q-2$,
\[
\Delta_{m\to n}^{(i)} = \ln \frac{\tau_{m\to n}(0)}{\tau_{m\to n}(H_{mn}\zeta^i)}.
\]
\State \textbf{Vertical step:} For $n \in [N]$ and $m\in\mathcal N(n)$, compute
\[
\Gamma_{n\to m}^{(i)} = \Lambda_n^{(i)} + \sum_{\substack{m'\in\mathcal N(n)\setminus m \\ \vphantom{j = tr(H_{m'n}\zeta^i)}}} \Delta_{m'\to n}^{(i)}.
\]
\State \textbf{Hard-decision step:} For $n \in [N]$, compute
\[
\hat\Lambda_n^{(i)} = \Lambda_n^{(i)} + \sum_{\substack{m\in\mathcal N(n) \\ \vphantom{j = tr(H_{mn}\zeta^i)}}} \Delta_{m\to n}^{(i)},
\]
\quad\, and set
\[
\hat e_n = \begin{cases} 0, & \hat\Lambda_n^{(i)} > 0\ \ \forall i, \\ \zeta^{\arg\min_i \hat\Lambda_n^{(i)}}, & \text{otherwise.} \end{cases}
\]
\Until{$\hat e \cdot H^T = \tilde s$ \textbf{or} $T_{\max}$ iterations are reached.}
\end{algorithmic}
\end{algorithm}

Note that the decoder proposed in~\cite{lai2021log} was intended to operate over $GF(4)$ and decode both $X$ and $Z$ type errors simultaneously for a qubit error correcting code. In general, one can use a decoder operating over $GF(q^2)$ to simultaneously decode both types of errors in a qudit code over $GF(q)$. However, for large $q$, the variable node operation over $GF(q^2)$ can become prohibitively complex and therefore, we have chosen in this work, to limit ourselves to the consideration of decoding a single type of error with the complementary error type being separately decoded by an identical decoder with similar performance.

When dealing with partially or fully stacked qudit codes, the decoder outlined in Algorithm~\ref{algo:refined_bp} has some important limitations. Since the trace function used in additive codes is a many-to-one function, the belief quantization step necessarily involves a certain amount of information loss that leads to a degradation in decoder performance. Moreover, the stacking operation typically adds a significant number of 4-cycles to the associated Tanner graph which can be detrimental to the performance of belief propagation~\cite{raveendran2014analysis}. For an order-$s$ stacked code, each check node of the base graph is repeated $s$ times. Since each of these $s$ repeated instances has an identical neighborhood, this leads to a very drastic increase in the number of 4-cycles as the number of stacks increases. The detrimental effect of these 4-cycles is partially countered by the fact that the messages being passed are non-binary which tend to have a longer effective oscillation period before messages begin repeating in a cycle. Nevertheless, countering the effect of these 4-cycles has the potential to significantly boost performance. In the next subsection, we describe an alternative strategy involving fully stacked codes that can counter this problem.

\subsection{Decoding Linear Codes}
\label{sec:decoder:linear}
Fully stacked (linear) codes offer a natural way to resolve the field mismatch problem between check and variable nodes via the use of non-binary syndromes rather than binary ones. Specifically, the use of the merge operator defined in Def.~\ref{def:merge-op} allows the usage of a standard classical non-binary decoder~\cite{zhang2016vlsi}. For these fully stacked constructions, the parity check matrices $H_x$ and $H_z$ can be written in the form,
\begin{equation}
    H_x = \begin{bmatrix}
        \alpha_0 H_{x0} \\
        \alpha_1 H_{x0} \\
        \vdots \\
        \alpha_{m-1} H_{x0}
    \end{bmatrix}; \qquad H_z = \begin{bmatrix}
        \beta_0 H_{z0} \\
        \beta_1 H_{z0} \\
        \vdots \\
        \beta_{m-1} H_{z0}
    \end{bmatrix},
\end{equation}
where $B_\alpha = \{\alpha_0, \alpha_1, \dots \alpha_{m-1}\}$ and $B_\beta = \{\beta_0, \beta_1, \dots \beta_{m-1}\}$ are a pair of primal and dual bases for $GF(q)$. From a graphical standpoint, each block $\alpha_i H_{x0}$ (or equivalently $\beta_i H_{z0}$) has an identical structure for each $i$ and therefore, the Tanner graph for the fully stacked code will likely contain a large number of cycles. To mitigate this issue, we can use a classical non-binary decoder on the unstacked graph $H_{x0}$ rather than the full graph $H_x$. To see this, consider an error pattern $e_z$ that yields a syndrome $s_z = tr(e_z H_x^T)$.
\begin{equation}
    s_z = tr(e_z H_x^T)
    \implies s_{z,i} = tr(e_z \cdot (\alpha_i H_{x0})^T),
\end{equation}
where $s_{z,i}$ is simply the syndrome corresponding to the $i^{th}$ block $\alpha_i H_{x0}$. Hence, we have simply partitioned the syndrome into $m$ blocks where each block corresponds to the syndrome obtained from a single block of the parity check matrix of the form $\alpha_i H_{x0}$. Then we can simply use the merge operator defined in Def.~\ref{def:merge-op}, with respect to the dual basis $B_\beta$, and apply Eq.~\ref{eq:basis-components} to get,
\begin{equation}
    \label{eq:merge-syn}
    \tilde{s}_z = \sum_{i=0}^{m-1} s_{z,i} \beta_i = \sum_{i=0}^{m-1} tr(\alpha_i e_z \cdot H_{x0}^T) \beta_i = e_z \cdot H_{x0}^T.
\end{equation}
Hence, the merged syndrome $\tilde{s}_z$ is now a non-binary value that can be used as a non-binary syndrome input to a standard classical non-binary decoder equipped with the check matrix $H_{x0}$. By using the check matrix $H_{x0}$ instead of $H_x$, we get rid of the identically structured blocks mentioned above, which is expected to yield a better decoding performance. It is worth noting, however, that this decoding approach can only be applied to fully stacked qudit codes. A similar analysis holds for $X$ errors in the context of the check matrix $H_{z0}$. Note that the above syndrome merge procedure requires the use of the dual basis $B_\beta$ for $Z$ errors and the primal basis $B_\alpha$ for $X$ errors. The full set of operations required for this approach is shown in Algorithm~\ref{algo:linear_bp}.

\section{Numerical Benchmarks}
\label{sec:results}
We now outline the results of some numerical experiments that we have performed for benchmarking concrete examples of the code constructions outlined above. We will begin by introducing a set of three candidate codes from the qudit hypergraph product and lifted product classes that we have used for numerical benchmarking. Subsequently, we shall discuss the performance of these candidate codes under the relevant decoding strategies. Finally, we shall discuss failure configurations for these message-passing decoders and provide a qualitative argument of why a non-binary error correction system performs better than its qubit counterpart.

\subsection{Prototype Code Constructions}
\label{sec:results:prototype-codes}
To benchmark the performance of the decoders discussed in the previous section, we perform a series of Monte Carlo experiments on a set of candidate codes designed using the constructions described earlier. Specifically, we focus on three codes, which we label as codes $A, B$ and $C$. Codes $A$ and $B$ belong to the hypergraph product class whereas code $C$ is a qudit lifted product code.
\begin{figure*}[t]
    \centering
    \includegraphics[width=\textwidth]{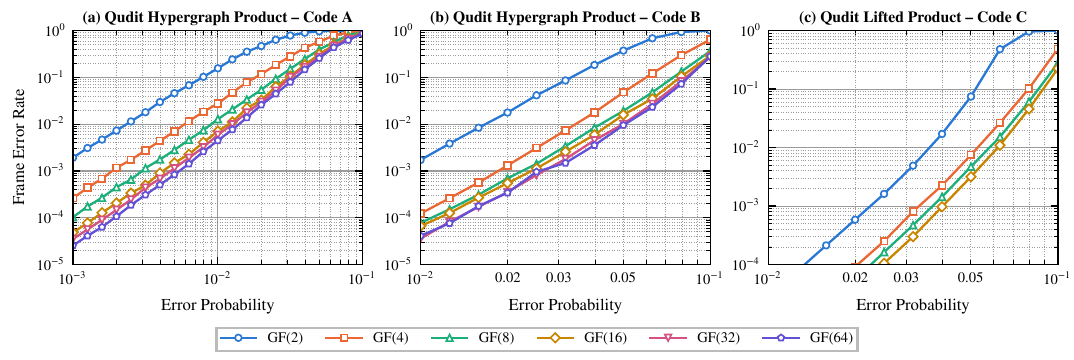}
    \caption{FER performance of fully stacked qudit QLDPC codes over GF($q$)
    under the BP decoder of Algorithm~1, for
    (a) the hypergraph product code $A$,
    (b) the hypergraph product code $B$, and
    (c) the lifted product code $C$.
    In all cases, the qudit codes ($q > 2$) achieve a lower FER than their qubit
    counterparts ($q = 2$). Each data point is a Monte Carlo estimate obtained by
    decoding $Z$-type errors with $H_X$, using up to $50$ BP iterations and running
    until $1000$ decoding failures are captured ($300$ for GF($32$) and GF($64$)
    in (b)).}
    \label{fig:rbp-sim}
\end{figure*}

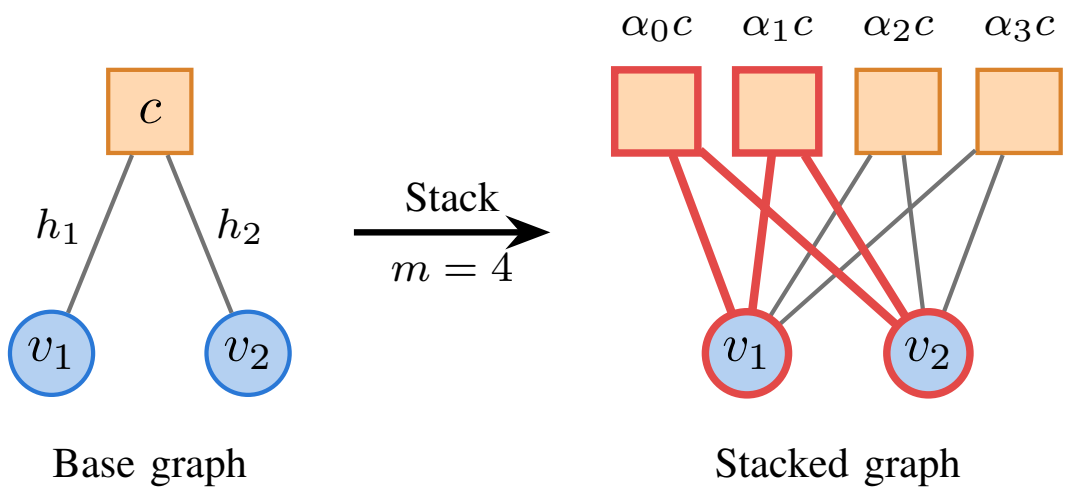
\begin{figure}[t]
    \centering
\definecolor{qVar}{HTML}{2A78D6}   
\definecolor{qChk}{HTML}{FF9933}   
\definecolor{qCyc}{HTML}{E34948}   
\begin{tikzpicture}[
    >=Stealth,
    vnode/.style={circle, draw=qVar, thick, fill=qVar!35, minimum size=5.5mm, inner sep=0pt},
    cnode/.style={rectangle, draw=qChk!85!black, thick, fill=qChk!38, minimum size=5.5mm, inner sep=0pt},
    edge/.style={thick, black!55},
    cyc/.style={line width=1.6pt, qCyc},
    lbl/.style={font=\footnotesize, text=black},
  ]

  \node[cnode] (c)  at (0, 1.6) {$c$};
  \node[vnode] (v1) at (-0.65, 0) {$v_1$};
  \node[vnode] (v2) at ( 0.65, 0) {$v_2$};
  \draw[edge] (c) -- node[lbl, left,  pos=0.45] {$h_1$} (v1);
  \draw[edge] (c) -- node[lbl, right, pos=0.45] {$h_2$} (v2);
  \node[lbl, align=center] at (0, -0.75) {Base graph};

  \draw[->, very thick] (1.35, 0.8) -- node[lbl, above, align=center] {Stack}
                                       node[lbl, below] {$m = 4$} (2.65, 0.8);

  \foreach \i/\x in {0/3.35, 1/4.15, 2/4.95, 3/5.75} {
    \node[cnode] (s\i) at (\x, 1.6) {};
    \node[lbl, above=0.5mm of s\i] {$\alpha_{\i} c$};
  }
  \node[vnode] (w1) at (3.95, 0) {$v_1$};
  \node[vnode] (w2) at (5.15, 0) {$v_2$};

  \foreach \i in {0,...,3} {
    \draw[edge] (s\i) -- (w1);
    \draw[edge] (s\i) -- (w2);
  }
  \draw[cyc] (s0) -- (w1) -- (s1) -- (w2) -- (s0);
  \foreach \n in {s0, s1} \node[cnode, draw=qCyc, line width=1.4pt] at (\n) {};
  \foreach \n/\t in {w1/$v_1$, w2/$v_2$}
    \node[vnode, draw=qCyc, line width=1.4pt] at (\n) {\t};

  \node[lbl, align=center] at (4.55, -0.75) {Stacked graph};
\end{tikzpicture}
    \caption{Stacking a check node of degree 2 with $m = 4$. The replicas
    $\alpha_i c$ share the same neighbors, so every pair of them forms a
    4-cycle (one shown in red), adding $^mC_2$ 4-cycles per check
    node for a degree-2 check node. This number is even higher for check nodes of degree $>2$.}
    \vspace{-\baselineskip}
    \label{fig:stacking-cycles}
\end{figure}

\begin{figure*}[t]
    \centering
    \includegraphics[width=\textwidth]{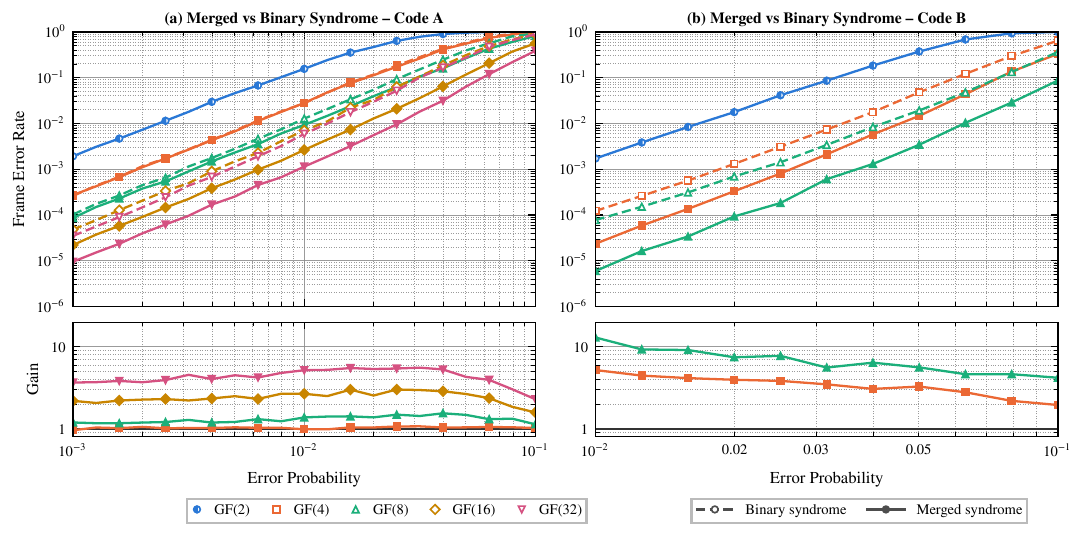}
    \caption{FER performance of the merged syndrome decoder (Algorithm~2; solid
    lines, filled markers) versus the binary syndrome decoder (Algorithm~1;
    dashed lines, open markers) for the fully stacked hypergraph product codes
    (a) $A$ and (b) $B$ over GF($q$). For $q = 2$ the two decoders coincide, so a
    single curve is shown. The lower panels show the gain
    $\mathrm{FER}_{\mathrm{binary}} / \mathrm{FER}_{\mathrm{merged}}$, where
    values above $1$ indicate an improvement from the merged syndrome. The gain
    increases with the field size $q$, reaching roughly an order of magnitude for
    code $B$ over GF($8$). Each data point is a Monte Carlo estimate obtained by
    running until $1000$ decoding failures are captured ($300$ for the merged
    syndrome decoder over GF($4$) and GF($8$) in (b)).}
    \label{fig:dec-sim}
\end{figure*}

For code $A$, we begin with two $8 \times 16$ quasi-cyclic binary check matrices $H_1, H_2$, where each matrix is built out of two $8 \times 8$ circulant blocks and both matrices have girth $8$. We replace each of their non-zero elements randomly with an arbitrary element in $GF(q)$, where $q$ is the qudit dimension under investigation and subsequently apply the order-$s$ stacked hypergraph product construction. The resulting code has $320$ variable nodes and $128s$ check nodes, where $s$ is the number of stacks used. For code $B$, we follow a similar approach but use a binary check matrix $H_1 = H_2 = H$ of size $18 \times 24$ obtained from \cite{roffe2020decoding} and \cite{raveendran2021trapping}. This code has a total of $900$ variable nodes and $432s$ check nodes, with the code parameters for the fully stacked case being $[[900, 36, 10]]$. Finally, code $C$ is obtained from a lifted product construction applied to the well-known Tanner code~\cite{tanner2003recursive}. Specifically, we start with the base matrix of the Tanner code and randomly replace the coefficient of each nonzero entry with an element of $GF(q)$ to get a base matrix over the $\F$-algebra $R_\ell$. Subsequently, we apply the construction described in Def.~\ref{def:lp}, with both $H_1$ and $H_2$ equal to the resulting base matrix, to construct code $C$. This code has parameters $[[1054, 140, 20]]$ in the fully stacked case, the number of logical qudits following from Cor.~\ref{cor:rate-css}. Since the lift size is $\ell = 31$, code $C$ is quasi-cyclic with circulant size $31$, and is thus an instance of the quasi-cyclic codes that motivate this work. Note, however, that unlike the family of Sec.~\ref{sec:good-lp}, where the second factor is a single polynomial $b \in R_\ell$, both factors of code $C$ are matrices over $R_\ell$. 

\subsection{Enhanced Decoding Performance with Qudits}
\label{sec:results:qudit-advantage}
To benchmark the performance of this approach, we perform a series of Monte Carlo experiments to estimate Frame Error Rates under the depolarizing noise model defined in Def.~\ref{def:depol-model}. For simplicity, we simply characterize the frame error rate performance of decoding for $Z$ errors using the check matrix $H_x$ and assume that the complementary error type will exhibit a similar performance. For all experiments performed in this work, we use up to $50$ iterations of belief propagation and run Monte Carlo trials as long as it takes to capture a total of $1000$ decoding failures (For some larger codes in higher fields, we reduce this number to $300$ to maintain a tractable simulation time). In principle, these decoding failures can occur either due to a convergence failure or due to a logical error. Logical errors are characterized by identifying the post-decoding residual error pattern $e_R = e - \hat{e}$, where $e, \hat{e}$ are the actual and estimated errors respectively, and subsequently testing whether $tr(e_R L_X^T) = 0$. Here $L_X$ is the $X$-logical matrix, whose rows correspond to a set of representatives for the logical $X$ operators. Similarly, the logical $Z$ operators can be defined using a similar matrix $L_Z$. The matrices $L_X$ and $L_Z$ for a given pair $(H_x, H_z)$ can be obtained using either Gottesman's~\cite{gottesman1997stabilizer} or Wilde's~\cite{wilde2009logical} algorithm for prime fields in conjunction with the split and merge transformations defined in Def.~\ref{def:split-op} and Def.~\ref{def:merge-op} respectively via the use of Thm.~\ref{thm:equiv}. It is worth noting, however, that most of the observed decoding failures in all experiments described henceforth, correspond to convergence failures rather than logical errors, demonstrating that decoder convergence is the dominant mode of failure.

Fig.~\ref{fig:rbp-sim} shows the frame error rate (FER) performance of the prototype code constructions described in Sec.~\ref{sec:results:prototype-codes} under the use of the decoder outlined in Algorithm~\ref{algo:refined_bp}. As can be seen from all three sets of simulations, decoding in higher fields provides a consistent advantage over binary decoding under the depolarizing noise model described in Def.~\ref{def:depol-model}. Indeed, all three plots show gains in the FER of 1--2 orders of magnitude when moving from $GF(2)$ to $GF(64)$ in powers of $2$. An interesting observation from Fig.~\ref{fig:rbp-sim}, however, is that this FER improvement seems to saturate out at higher fields. For example, while the jump in performance from $GF(2)$ to $GF(4)$ is fairly large, this gain progressively becomes smaller at higher field sizes. We attribute this behavior to the stacking operation used in the construction of these codes. In particular, Fig.~\ref{fig:stacking-cycles} shows how the stacking operation can introduce large amounts of 4-cycles in the Tanner graph of the code. Specifically, Fig.~\ref{fig:stacking-cycles} shows that a degree-2 check node can introduce up to $\binom{m}{2}$ 4-cycles when stacked $m$ times and this number of cycles introduced is even higher for higher degree check nodes. In Sec.~\ref{sec:results:four-cycles}, we characterize which 4-cycles are harmful to decoding and return to the effect of these stacking-induced cycles.

\subsection{Merged vs Binary Syndrome Decoding}
\label{sec:results:decoder-compare}
We now proceed to compare the performance of the two decoding approaches described in Sec.~\ref{sec:decoder} for fully stacked constructions. As discussed in Sec.~\ref{sec:decoder:linear}, the merged syndrome decoder described in Algorithm~\ref{algo:linear_bp} operates on the base Tanner graph rather than the stacked graph. Hence, it is able to circumvent the 4-cycles described in Fig.~\ref{fig:stacking-cycles}. Fig.~\ref{fig:dec-sim} shows a comparison of these two decoding approaches for the fully stacked hypergraph product codes described in Sec.~\ref{sec:results:prototype-codes}. As expected, the lower number of 4-cycles allows the merged syndrome decoder to achieve a significantly lower FER than the corresponding binary syndrome decoder with the same field size. On the lower part of Fig.~\ref{fig:dec-sim}, we also plot the FER gain obtained by using the merged syndrome decoder over the binary syndrome decoder. This gain is defined simply as $\mathrm{FER}_{\mathrm{binary}} / \mathrm{FER}_{\mathrm{merged}}$. We observe that the FER gain increases with an increasing field size, which can be attributed to the fact that the number of 4-cycles introduced via stacking is greater for higher field sizes, thus contributing to a greater FER loss in higher field sizes for the binary syndrome decoder. Consequently, the FER saturation effect observed in the binary syndrome decoder for higher field sizes, is significantly less pronounced for the merged syndrome decoder. We revisit the source of this gain in Sec.~\ref{sec:results:four-cycles}, after characterizing which 4-cycles are harmful to a non-binary decoder.
\begin{figure*}[t]
    \centering
    \includegraphics[width=\textwidth]{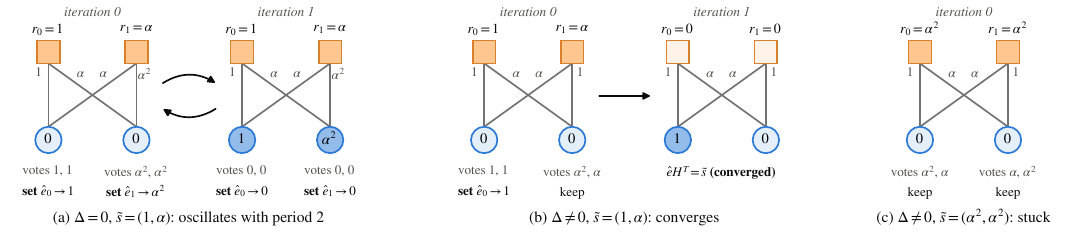}
    \caption{Syndrome-based symbol-flipping decoding on a single 4-cycle over
    GF($4$), with $\alpha^2 = \alpha + 1$. Checks show the residual syndrome
    $r_m$ and variables the estimate $\hat{e}_n$ (dark: nonzero); edges show
    $h_{mn}$. A symbol is updated only when both of its checks vote for the same
    value. (a) $\Delta = 0$: decoder oscillates between $\hat{e} = 0$ and
    $\hat{e} = (1, \alpha^2) \in \ker H$. (b) $\Delta \neq 0$: converges in
    one iteration. (c) $\Delta \neq 0$, weight-2 error: decoder is stuck 
    due to split votes.}
    \label{fig:sf-mechanism}
\end{figure*}

\subsection{Understanding Qudit Advantage}
\label{sec:results:four-cycles}
Traditionally, an analysis of the failure configurations of belief propagation (BP) variants for qubit decoding is done via an identification of trapping sets for a simpler discrete decoder, such as a bit-flipping decoder~\cite{raveendran2021trapping}. While the dynamics of BP and bit-flipping decoders are, in general, different, it is known that failure configurations for the bit-flipping decoder often fail to converge for the BP decoder as well. Following a similar philosophy, and with the understanding that the presence of 4-cycles is a dominant source of decoding failures for BP, we shall attempt to provide an intuitive explanation for the results in the previous subsections by exploring a simple toy model, consisting of a symbol flipping decoder applied to a non-binary 4-cycle. For simplicity, we shall restrict our analysis to the merged syndrome decoder of Algorithm~\ref{algo:linear_bp}. We begin with a brief description of the message passing decoder.

Similar to the BP decoder, the symbol-flipping decoder operates via the iterative use of check and variable-node update rules. Each variable node maintains a state $\in GF(q)$, corresponding to the error at that variable node, initialized to zero. In the initialization phase, this state is passed to the checks as the outgoing messages. In the check node update rule, we simply solve for the outgoing message on each edge as a function of the incoming messages from all other edges incident on the same check node. Specifically,
\begin{equation}
    \Delta_{m\to n} = H_{mn}^{-1}\Big(\tilde{s}_m
        - \sum_{n'\in\mathcal{N}(m)\setminus n} H_{mn'}(\Gamma_{n'\to m})\Big),
    \label{eq:sf-check-update}
\end{equation}
where $\tilde{s}$ is the merged syndrome defined in Eq.~\ref{eq:merge-syn}. Similarly, at the variable node, we compare all incoming messages and update the current state of the decoder if the number of messages instructing a flip to a specific element of $GF(q)$ exceeds a certain threshold. This threshold is a tunable parameter of the decoder in practice, but for the analysis of 4-cycles, since only two incoming messages are available, we simply choose to flip the state if both incoming messages agree on the choice of flip and keep the current state otherwise. The updated state is then sent to the check nodes as the outgoing messages and this process is iterated until convergence or timeout.
\begin{figure}[t]
    \centering
    \includegraphics[width=\linewidth]{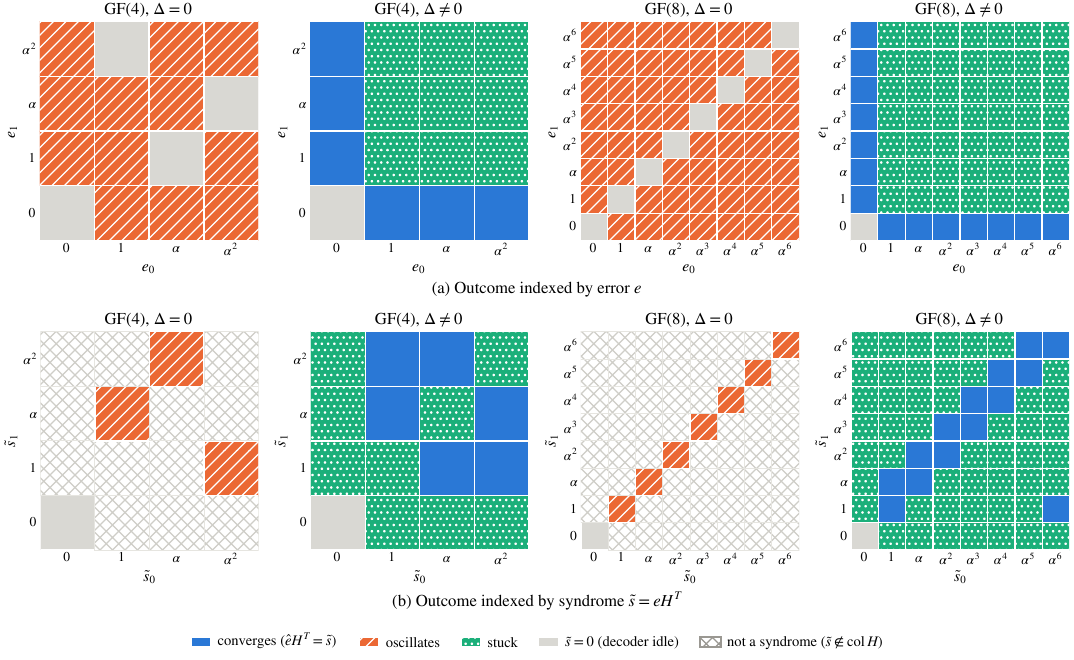}
    \caption{Outcome of syndrome-based symbol-flipping decoding on a single
    4-cycle over GF($4$) and GF($8$), indexed by (a) the error $e$ and (b) the
    syndrome $\tilde{s} = eH^T$. For $\Delta = 0$, every nonzero syndrome leads to
    oscillation. For $\Delta \neq 0$, the decoder converges only on the
    syndromes of weight-1 errors and is stuck on all others.}
    \label{fig:sf-outcome-maps}
\end{figure}

\begin{figure*}[t]
    \centering
    \includegraphics[width=\textwidth]{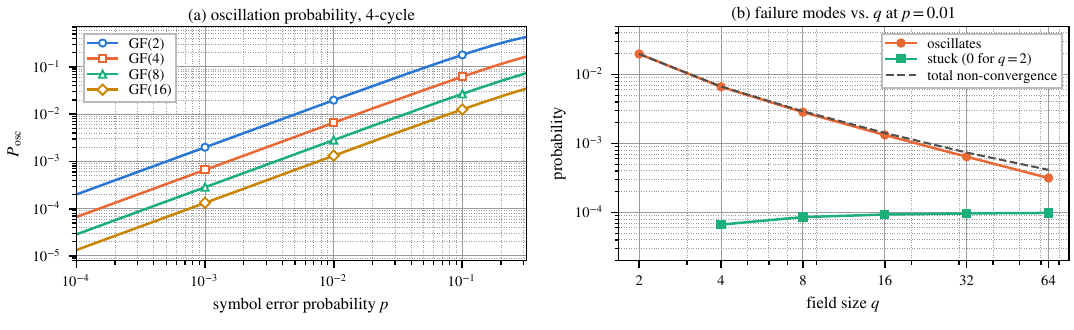}
    \caption{Failure probabilities of syndrome-based symbol-flipping decoding
    on a single 4-cycle over GF($q$) with uniformly random nonzero edge labels,
    computed exactly by exhaustive enumeration.
    (a) Oscillation probability versus symbol error probability $p$.
    (b) Probability of oscillating, of being stuck, and of not converging
    versus $q$ at $p = 0.01$. Oscillation requires $\Delta = 0$, which occurs
    with probability $1/(q-1)$.}
    \label{fig:sf-osc-prob}
\end{figure*}

\begin{definition}
    \label{def:cycle-det}
    Consider a 4-cycle over $GF(q)$ with $q = p^m$ being some prime power. Let us represent the 4-cycle using a parity check matrix,
    \begin{equation}
        \label{eq:cycle-matrix}
        H = \begin{bmatrix}
            h_{00} & h_{01} \\
            h_{10} & h_{11}
        \end{bmatrix}.
    \end{equation}
    We define the cycle determinant $\Delta$ of the 4-cycle as the determinant of its associated parity check matrix, namely,
    \begin{equation}
        \Delta = h_{00} h_{11} - h_{01} h_{10}.
    \end{equation}
\end{definition}

Fig.~\ref{fig:sf-mechanism} shows the possible dynamics of the symbol-flipping decoder applied to a single 4-cycle over $GF(4)$. Depending on the cycle determinant $\Delta$, three possibilities can occur: (a) the state of the variable nodes oscillates (Fig.~\ref{fig:sf-mechanism}(a)), (b) the decoder converges (Fig.~\ref{fig:sf-mechanism}(b)), and (c) the decoder gets stuck due to split votes (Fig.~\ref{fig:sf-mechanism}(c)). In general, decoder convergence does not guarantee correct decoding, since a logical error may still occur, but as mentioned in Sec.~\ref{sec:results:qudit-advantage}, convergence failure is the dominant failure mode. Hence, we restrict our attention to the consideration of decoder convergence. Among the three possible scenarios, note that the third scenario of split votes is specific to the symbol flipping decoder and does not always arise in the case of BP since the messages passed in BP consist of probabilistic information in the form of LLRs rather than hard decisions. Hence, we may infer that scenario (a), corresponding to decoder state oscillation, is the most prominent mode of decoder failure due to 4-cycles.

In Fig.~\ref{fig:sf-outcome-maps}, we enumerate all possible error patterns for 4-cycles in $GF(4)$ and $GF(8)$ and characterize their behavior according to the three scenarios of Fig.~\ref{fig:sf-mechanism}. In general, this behavior depends on the specific error pattern that occurs but also on the parity check matrix associated with the cycle. We observe, however, that the effect of the parity check matrix on the decoder dynamics can be completely captured by whether the cycle determinant $\Delta$ is zero or not. Hence, in Fig.~\ref{fig:sf-outcome-maps}, we exhibit two cases for each of $GF(4)$ and $GF(8)$ depending on whether $\Delta = 0$ or not. Fig.~\ref{fig:sf-outcome-maps}(a) shows the outcomes indexed by the error pattern and Fig.~\ref{fig:sf-outcome-maps}(b) shows the same indexed by the syndrome. Note that not all syndromes are valid (some do not lie in the column space of the cycle check matrix) and hence, Fig.~\ref{fig:sf-outcome-maps}(b) has certain entries marked as not possible. The dependence of the decoder dynamics on the cycle determinant $\Delta$ can thus be characterized by the following theorem.
\begin{theorem}
\label{thm:cycle-det}
    Consider a symbol-flipping decoder applied to a 4-cycle over $GF(q)$, with $q = p^m$ a prime power, with error pattern $e$ and merged syndrome $\tilde{s} = eH^T$. Let $\Delta$ be its cycle determinant as defined in Def.~\ref{def:cycle-det}. Then,
    \begin{enumerate}
        \item If $\Delta = 0$ and $e \notin \ker H$ (equivalently, $\tilde{s} \neq 0$), the decoder never converges. Its state oscillates with
        period $2$.
        \item If $\Delta \neq 0$ and $\wt{e} = 1$, it
        converges in a single iteration with $\hat{e}^{(1)} = e$. 
        \item If $\Delta \neq 0$ and $\wt{e} = 2$, then
        $\hat{e}^{(t)} = 0$ for all $t$, and the decoder is stuck at a fixed point with
        $\hat{e}^{(t)} H^T \neq \tilde{s}$.
    \end{enumerate}
\end{theorem}
\begin{proof}
    Let $\hat{e}^{(t)}$ denote the decoder state at iteration $t$ and $r^{(t)} = \tilde{s} - \hat{e}^{(t)}H^T$ denote the residual syndrome at iteration $t$. Let $w_n = (h_{0n}, h_{1n})$ denote the $n$th column of $H$, so that $\tilde{s} = e_0w_0 + e_1w_1$. Substituting $\Gamma_{n'\to m} = \hat{e}^{(t)}_{n'}$ into Eq.~\ref{eq:sf-check-update},
    \begin{equation}
        \Delta_{m\to n}
        = h_{mn}^{-1}\Big(r^{(t)}_m + h_{mn}\hat{e}^{(t)}_n\Big)
        = \hat{e}^{(t)}_n + h_{mn}^{-1} r^{(t)}_m .
        \label{eq:sf-vote-residual}
    \end{equation}
    As $\hat{e}^{(0)} = 0$, we have $r^{(0)} = \tilde{s}$ and $\Delta_{m\to n} = h_{mn}^{-1}\tilde{s}_m^{\vphantom{1}}$. Hence variable $n$ is updated to a nonzero value $a$ if and only if $h_{0n}^{-1}\tilde{s}_0^{\vphantom{1}} = h_{1n}^{-1}\tilde{s}_1^{\vphantom{1}} = a \neq 0$, that is, if and only if $\tilde{s} = a\,w_n$ for some $a \neq 0$.
 
    \emph{Case $\Delta = 0$.} The columns $w_0$ and $w_1$ are linearly dependent, and since all $h_{mn} \neq 0$, the column space of $H$ is the line spanned by either column. As $\tilde{s} = eH^T$ is a nonzero element of this space, $\tilde{s} = a_nw_n$ with $a_n = h_{0n}^{-1}\tilde{s}_0^{\vphantom{1}} \neq 0$ for both $n = 0, 1$. Therefore both variables are updated, and $\hat{e}^{(1)} = (a_0, a_1) \neq 0$. Then $\hat{e}^{(1)}H^T = a_0w_0 + a_1w_1 = 2\tilde{s}$, so $r^{(1)} = -\tilde{s} \neq 0$ and the decoder has not converged. By Eq.~\ref{eq:sf-vote-residual}, $\Delta_{m\to n} = a_n - h_{mn}^{-1}\tilde{s}_m^{\vphantom{1}} = a_n - a_n = 0$ for both checks $m$, so both variables are reset and $\hat{e}^{(2)} = 0 = \hat{e}^{(0)}$. Since the update rule depends only on the current state, the sequence repeats with period $2$. Neither state satisfies the syndrome, because $r^{(0)} = \tilde{s} \neq 0$ and $r^{(1)} = -\tilde{s} \neq 0$.
 
    \emph{Case $\Delta \neq 0$.} The columns $w_0$ and $w_1$ are linearly independent, so $\tilde{s}$ is a nonzero multiple of at most one of them, and at most one variable is updated at $t = 0$. Moreover, $\tilde{s} = e_0w_0 + e_1w_1 = a\,w_n$ holds if and only if $e$ has weight one with $e_n = a$. If $\wt{e} = 1$ with support $\{n\}$, only variable $n$ is updated, to $e_n$. Then $\hat{e}^{(1)} = e$ and $r^{(1)} = 0$, so the decoder converges. If $\wt{e} = 2$, no variable is updated, so $\hat{e}^{(1)} = \hat{e}^{(0)} = 0$. The decoder therefore remains at $\hat{e} = 0$ with $r = \tilde{s} \neq 0$ for all subsequent iterations. In neither case is a state revisited after a change, so no oscillation occurs.
\end{proof}

For the case of binary decoders, the cycle check matrix of Eq.~\ref{eq:cycle-matrix} can only exhibit entries identical to $1$. Hence, the cycle determinant is always identical to 0 and consequently, 4-cycles always lead to decoder oscillations. In the non-binary case, however, not all 4-cycles are harmful to the decoder and oscillations occur only when the corresponding cycle determinant is zero. With the entries in the cycle check matrix drawn uniformly at random from $GF(q) \setminus \{0\}$, the probability that the cycle determinant is zero is given by $1/(q - 1)$ and hence, the probability that any given cycle will induce oscillations decreases with increasing field size. This offers an explanation for the performance improvement of the merged syndrome decoder seen in Fig.~\ref{fig:dec-sim} at higher field sizes; the same mechanism is expected to contribute to the trend in Fig.~\ref{fig:rbp-sim}, although the check-node operation of Algorithm~\ref{algo:refined_bp} differs from Eq.~\ref{eq:sf-check-update}. Indeed, for the simple case of the toy model of a 4-cycle that we have considered in this section, we can formalize the probability of oscillation under a depolarizing noise channel as follows.
\begin{corollary}
    \label{cor:osc-prob}
    Let the entries of $H$ in Def.~\ref{def:cycle-det} be drawn independently and uniformly from $GF(q)\setminus\{0\}$, and let each $e_n$ follow the depolarizing channel of Def.~\ref{def:depol-model}, i.e., independently $e_n = 0$ with probability $1-p$ and each nonzero value with probability $p/(q-1)$, where $p$ is given by Eq.~\ref{eq:depol-prob}. Then the symbol-flipping decoder oscillates with probability
    \begin{equation}
        P_{\mathrm{osc}} = \frac{1}{q-1}\left(2p - p^2 - \frac{p^2}{q-1}\right).
        \label{eq:sf-osc-prob}
    \end{equation}
\end{corollary}
 
\begin{proof}
    For any fixed $h_{00}, h_{01}, h_{10}$, exactly one value of $h_{11}$ gives $\Delta = 0$, so $\Pr[\Delta = 0] = 1/(q-1)$. By Thm.~\ref{thm:cycle-det}, the decoder oscillates if and only if $\Delta = 0$ and $e \notin \ker H$. When $\Delta = 0$, $\ker H$ consists of $0$ and the $q-1$ weight-2 vectors $(x, -h_{00}^{-1}h_{01}^{\vphantom{1}}x)$, $x \neq 0$. Hence $\Pr[e \notin \ker H] = 1 - (1-p)^2 - (q-1)\,p^2/(q-1)^2$, and multiplying by $\Pr[\Delta = 0]$ gives the above result.
\end{proof}

We plot this probability of oscillations, as described by Eq.~\ref{eq:sf-osc-prob}, in Fig.~\ref{fig:sf-osc-prob}(a) for various field sizes in powers of $2$. As observed in Fig.~\ref{fig:sf-osc-prob}(a), there is a clear improvement in oscillation probability with increasing field sizes. Indeed, this performance improvement is very similar to the trends observed in the Monte Carlo simulation results of Fig.~\ref{fig:rbp-sim} and Fig.~\ref{fig:dec-sim}. This provides strong evidence of the fact that the source of the observed performance improvement is indeed the fact that not all 4-cycles in a non-binary field are harmful for the decoder. In Fig.~\ref{fig:sf-osc-prob}(b), we fix the depolarizing probability $p$ to $0.01$ and plot the oscillation probability, the total non-convergence probability and the probability of the decoder getting stuck as a function of field size. As observed in Fig.~\ref{fig:sf-osc-prob}(b), the contribution of the cases where the decoder gets stuck to the total non-convergence probability is negligible compared to the contribution from the oscillation probability. This means that the oscillation probability of Fig.~\ref{fig:sf-osc-prob}(a) is a sufficiently good approximation of the total non-convergence probability. Moreover, in the case of BP, which operates with soft probabilistic information rather than the hard decisions of the symbol flipping decoder, the decoder does not necessarily get stuck in all the cases that it does for the case of our simple toy model. To summarize, cycle determinants for 4-cycles are always zero for binary decoders but are potentially non-zero for non-binary decoders, which essentially contributes to an improved FER performance for non-binary decoders. Moreover, the probability of a non-zero cycle determinant increases with increasing field sizes and therefore, larger field sizes exhibit lower FERs.

At this point, it is worth noting that the use of non-zero cycle determinants is indeed a well-known design criterion for the design of classical non-binary LDPC codes~\cite{poulliat2008design, peng2007design}. Indeed, Thm.~\ref{thm:cycle-det} is simply a description of similar ideas for the case of quantum decoders wherein non-binary syndrome decoding is required. The FER improvement offered by the merged syndrome decoder over the binary syndrome decoder, as shown in Fig.~\ref{fig:dec-sim} also has classical roots in the ideas such as generalized belief propagation~\cite{yedidia2000generalized}, wherein multiple nodes in the Tanner graph are merged into super-nodes with the goal of reducing cycles. Although Thm.~\ref{thm:cycle-det} is stated for the merged syndrome decoder, it also offers some intuition for why the stacking-induced 4-cycles of Fig.~\ref{fig:stacking-cycles} are particularly detrimental to the binary syndrome decoder. The replicas of a stacked check constrain the same $GF(q)$ linear combination of their neighbors, each through a different $GF(p)$ projection, and therefore provide no independent information to distinguish the variables involved, much like a 4-cycle with a vanishing cycle determinant. Unlike the 4-cycles of the base graph, these cycles do not become less harmful as the field size grows, and their number increases with the stacking order. This is consistent with the FER saturation in Fig.~\ref{fig:rbp-sim} and with the growing gain of the merged syndrome decoder in Fig.~\ref{fig:dec-sim}.

It should also be noted that the FER improvements demonstrated in this section are very specific to the depolarizing channel. In particular, the error probability in the x-axes of Fig.~\ref{fig:rbp-sim} and Fig.~\ref{fig:dec-sim} correspond to the probability $p$ as described in Eq.~\ref{eq:depol-prob} and there is a priori no reason to believe that a qubit depolarizing error channel with probability $p$ is comparable to a qudit depolarizing error channel with the same error probability. In a similar classical scenario, the available channel bandwidth would be the reference parameter and a fair comparison would express $p$ for various field sizes as a function of the available bandwidth. Indeed, these ideas have led to the development of the field of classical coded-modulation~\cite{anderson2003coded}, which involves a joint design of the modulation scheme in conjunction with the error correcting code to get a net coding gain. In the quantum case, however, the system bandwidth is no longer the relevant parameter of choice and a proper full system analysis would require knowledge of the relevant qudit noise models for the specific hardware modality (superconducting, photonic, trapped-ion etc.) realizing the qudit system as well as any bosonic code (i.e., modulation scheme) used~\cite{raveendran2022finite, borah2025fault}. For the purposes of this work however, we have ignored such hardware-specific constraints and have approximately treated all qudit channels uniformly with respect to their error probability $p$. We shall delegate the question of a full-fledged design of a quantum coded-modulation scheme as a topic of future research.

\section{Conclusions}
\label{sec:conclusion}
In this work, we have investigated the extension of qubit QLDPC codes to qudit systems through the use of non-binary codes on higher degree finite fields. Specifically, we have investigated the generalization of the hypergraph product family and the lifted product family to qudits of higher dimensions. For the hypergraph product we have provided upper and lower bounds on the encoding rate, and for the lifted product an exact expression for the number of logical qudits. For the qudit lifted product construction, we have shown that, in the fully stacked case, lifted Tanner codes on cyclic lifts of expander graphs yield families of quasi-cyclic qudit LDPC codes with $\Theta(\log N)$ logical qudits and an almost linear minimum distance $\Theta(N/\log N)$. Restricting to cyclic lifts trades the asymptotically good scaling achievable with non-abelian lifts for a quasi-cyclic structure that is well suited to hardware decoder implementations, which is particularly valuable for non-binary decoders. We have investigated the performance of both constructions using two decoding approaches, namely, the log-domain BP decoder of \cite{lai2021log} and a classical non-binary decoder with merged syndrome. For both cases, qudit codes offer an improved FER over their qubit counterparts under a depolarizing channel with the same symbol error probability, although, as discussed in Sec.~\ref{sec:results}, a fair comparison across field sizes ultimately depends on the noise model of the specific hardware platform. For fully stacked codes, the performance improvement is more pronounced for the merged syndrome approach. But this approach does not generalize to the partially stacked versions of the qudit constructions.

While we have demonstrated several advantages offered by qudit constructions in this work, several open questions are yet to be addressed. For instance, a formal bound on the minimum distance of the partially stacked constructions is not known. On the theoretical side, tightening the bounds on encoding rates for non-fully-stacked constructions, and improving the constants in the minimum distance bound of Theorem~\ref{thm:distance} for general polynomials $b$, would strengthen the framework considerably. A natural next step is to apply the framework developed here to the lifted product of two cyclically lifted Tanner codes, which, as observed by Panteleev and Kalachev~\cite{panteleev2022asymptotically}, yields quasi-cyclic codes with a linear number of logical qudits and minimum distance $\Theta(N/\log N)$. Similarly, extending the stacked construction to non-abelian group algebras would recover the asymptotically good codes of~\cite{panteleev2022asymptotically} in the fully stacked case and could shed light on the partially stacked case. From a decoding perspective, extending the merged syndrome approach to non-fully-stacked constructions is an important research direction. Finally, the classification of logical operators and Clifford operators for qudit QLDPC codes presents an important and interesting research question. 

\section*{Acknowledgment}
S.K. Borah, B. Vasi\'{c}, and N. Rengaswamy were generously supported by the U.S. Department of Energy, Office of Science, National Quantum Information Science Research Centers, Superconducting Quantum Materials and Systems Center (SQMS), under Contract No. 89243024CSC000002. V. Lalitha acknowledges the funding support from the National Quantum Mission, an initiative of the Department of Science and Technology, Govt. of India. She also thanks the support provided by the Foundation for QC Innovation (FQCI), DST-NQM T-Hub at IISc Bengaluru, in facilitating this project. V. Lalitha also acknowledges the support of ANRF Advanced Research Grant (ANRF/ARG/2025/009230/ENS) for this work.
Initial discussions for this collaboration
took place while V. Lalitha and N. Rengaswamy were visiting the Simons Institute for the Theory of Computing in Spring 2024.

The authors acknowledge the use of an AI model (Claude Opus 5.5) primarily for language editing, including improving sentence structure and readability. It was also used to fill some gaps in the proof of the minimum distance. AI assistance was also used in preparing some figure captions and developing auxiliary scripts for simulation management and data visualization, including automating parameter sweeps and generating plots from simulation results. The core implementation of the quantum error correction decoders, Monte Carlo simulations, and the underlying research methodology were developed by the authors. All AI-assisted outputs, including code, were reviewed, tested, and verified by the authors, who take full responsibility for the contents of this paper.

\bibliographystyle{IEEEtran}
\bibliography{IEEEabrv, references}

\end{document}